\documentclass[journal ]{new-aiaa}
\usepackage[utf8]{inputenc}
\usepackage{textcomp}

\usepackage{graphicx}
\usepackage{amsmath}
\usepackage[version=4]{mhchem}
\usepackage{siunitx}
\usepackage{longtable,tabularx}
\usepackage{epstopdf}
\usepackage{color}
\usepackage{url}
\usepackage{subcaption}
\usepackage{multirow}

\newcommand{\ignore}[1]{}

\newcommand{\norm}[1]{\left\Vert#1\right\Vert} 

\newcommand{\bbm}{\begin{bmatrix}}
\newcommand{\ebm}{\end{bmatrix}}
\newcommand{\bma}[1]{\left[\begin{array}{#1}}
\newcommand{\ema}{\end{array}\right]}

\DeclareMathAlphabet{\mbf}{OT1}{ptm}{b}{n}
\newcommand{\mbs}[1]{{\boldsymbol{#1}}}

\newcommand{\mbfhat}[1]{{\hat{\mbf{#1}}}}

\def\dotb{{\raisebox{-0.6ex}{ \kern0.2ex\raisebox{0.8ex}{\tiny $\circ$}}}}
\def\ddota{{\raisebox{-0.6ex}{ \raise0.2ex\hbox{ \LARGE $\cdot\hspace*{-0.2ex}\cdot$}}}}
\def\ddotb{{\raisebox{-0.6ex}{ \kern0.2ex\raisebox{0.8ex}{\tiny $\circ\circ$}}}}

\newcommand{\ura}[1]{{\underrightarrow{{#1}}}}

\newcommand{\trans}{{\ensuremath{\mathsf{T}}}} 
\newcommand{\beq}{\begin{equation}}
\newcommand{\eeq}{\end{equation}}
\newcommand{\bdis}{\begin{displaymath}}
\newcommand{\edis}{\end{displaymath}}
\newcommand{\beqarray}{\begin{eqnarray}}
\newcommand{\eeqarray}{\end{eqnarray}}
\newcommand{\beqarraynn}{\begin{eqnarray*}}
\newcommand{\eeqarraynn}{\end{eqnarray*}}

\newcommand{\utimes}{{\raisebox{-0.6ex}{ \kern-1.0ex\raisebox{0.6ex}{\small$\mathsf{v}$}}} } 

\title{Time-Varying Model Predictive Attitude Control for Magnetically Actuated Dual-Spin Satellites}

\author{Robert D. Halverson \footnote{Ph.D. Candidate, Department of Aerospace Engineering \& Mechanics, AIAA student member.} and Ryan J. Caverly \footnote{Assistant Professor, Department of Aerospace Engineering \& Mechanics, AIAA member.}}
\affil{University of Minnesota - Twin Cities, Minneapolis, MN, 55455}

\begin{document}

\maketitle

\begin{abstract}
Attitude control hardware for small satellites is often limited in power and space availability given the importance of the science instruments they exist to transport. To mitigate this, a dual-spin stabilized satellite actuated via magnetic torque rods reduces the space and power required of the attitude control system, but may require advanced control policies. This paper explores the attitude control of a magnetically actuated dual-spin stabilized CubeSat with model predictive control using time-varying prediction dynamics. An inertial pointing objective is used as a representative mission, where the satellite is able to deviate from its nominal orientation within some allowable amount. Three time-varying MPC policies are developed and compared to ensure the system does not violate constraints while minimizing control effort. These policies include a prediction model that accounts for the orbital position of the satellite and two iterative approaches that incorporate control inputs through either propagation of the linear dynamics, or nonlinear propagation with successive linearization. Results demonstrate that the nonlinear prediction policy outperforms other prediction methods with regards to not only minimal actuation, but to constraint satisfaction as well while imparting minimal computational burden. 
\end{abstract}

\section{Introduction}
\lettrine{T}{he} small satellite revolution is bringing forth a new age of space exploration technologies, causing spacecraft to become increasingly small while continuing to deliver important science payloads and instruments to low-Earth orbit (LEO) and beyond~\cite{sweeting2018modern}. Cube Satellites (CubeSats) are standardized and cost-effective platforms that are accessible to not only the spacecraft manufacturing industry, but many universities and research institutions alike~\cite{villela2019towards,poghosyan2017cubesat}. Science instruments often use a great amount of power and space within CubeSats, which limits the power available for use by other important components including attitude determination and control system (ADCS) hardware. The goal of the ADCS is to determine the satellite's orientation with respect to some inertial reference frame in real time while simultaneously controlling the satellite via available actuators. CubeSats are often equipped with three reaction wheels that are nominally non-spinning and exchange momentum with the satellite bus to produce a torque in a desired direction. Magnetic torque rods are usually then included to interact with the Earth's magnetic field when actuated to produce a small torque for the purpose of momentum management. 

This paper explores the problem where a CubeSat does not have enough space or power available for three reaction wheels. Rather, the spacecraft is equipped with one nominally spinning wheel (i.e., a `momentum wheel') and three magnetic torque rods mounted orthogonally within the spacecraft. The spacecraft is spun around the same axis of the wheel's momentum vector, which defines the dual-spin stabilized configuration~\cite{DeRuiter2013,hughes2012spacecraft}. This configuration provides passive stability properties such that a gyroscopic stiffness is provided to the spin axis. 

\begin{figure}[t!]
	\includegraphics[width=0.45\linewidth]{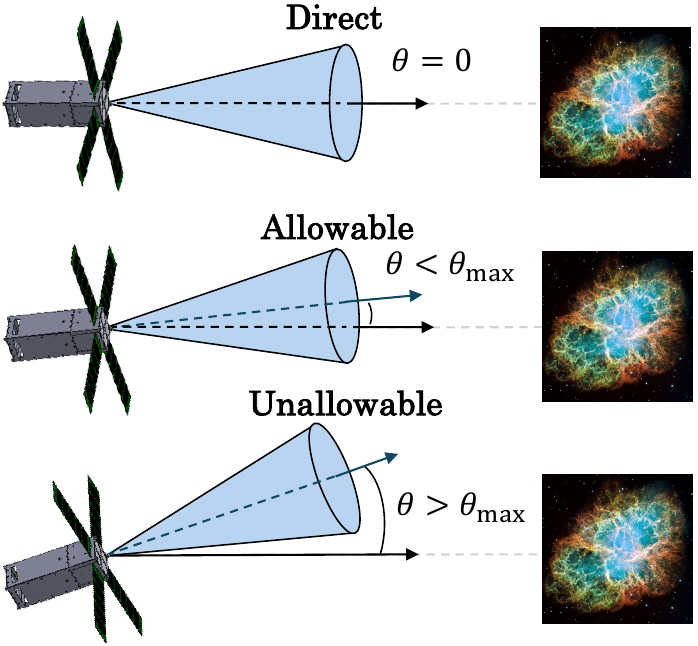} 
	\centering
	\caption{An illustration of three configurations for a 3U CubeSat pointing at the distant Crab Nebula.}\label{fig:DSS_pointing}
\end{figure}

This work is inspired by the EXACT (Experiments in X-ray Characterization and Timing) satellite being designed and built at the University of Minnesota~\cite{runnels2017signal}. This mission employs a 3U (30cm x 10cm x 10cm) CubeSat that gathers data from the HAFX (Hard and Fast X-Ray Spectrometer) science instrument. The primary objective of this mission requires the instrument boresight attached to the CubeSat to point at the Crab Pulsar within a specified half-cone angle. Figure~\ref{fig:DSS_pointing} illustrates the different pointing cone configurations for this representative mission. Importantly, the science instrument allows the CubeSat to spin about its boresight axis, where a minimum spin rate is required based on the novel attitude determination system that was developed for the satellite~\cite{laughlin2022applications}.

There is a considerable amount of literature regarding the use of solely magnetic actuation beginning with some of the first publications investigating magnetic actuation for spinning satellites in the 1960's~\cite{ergin1965magnetic}. A useful survey on the use of various magnetic control approaches for spacecraft including linear and nonlinear control theory, as well as predictive control is found in Ref.~\cite{silani2005magnetic}, while a deep-dive into magnetic attitude control for small satellites is presented in Ref.~\cite{ovchinnikov2019survey}. Particularly relevant literature in this area include an investigation into periodic linear quadratic control for nadir-pointing spacecraft in a bias momentum configuration using magnetic actuation~\cite{pittelkau1993optimal}, sliding-mode control for satellites in the presence of disturbance torques~\cite{zhou2017}, and an investigation into controllability and stability via linear quadratic regulation (LQR) with magnetic actuation~\cite{morozov2020satellite}. Other work includes the application of periodic LQR with proven steady-state errors of less than one degree in a nadir-pointing mission and robustness to model uncertainties and unmodeled dynamics~\cite{psiaki2001magnetic}, as well as the application of full-state feedback via LQR with an observer-controller design for magnetic attitude control~\cite{reyhanoglu2011three}.

Much of existing literature uses some form of LQR control for magnetic actuation, however this often requires extensive tuning and risks an aggressive and sometimes over-actuating control law. Further, the use of a static or pre-scheduled control gain introduces the risk of limited robustness to uncertainty in system and environment dynamics. To mitigate this, a predictive control policy can be utilized. Model predictive control (MPC) has been used for a myriad of aerospace control applications including autonomous station keeping of satellites~\cite{halverson2025,caverly2020electric}, spacecraft attitude control and relative motion control~\cite{di2018real,petersen2023safe}, and many others as outlined in the survey in Ref.~\cite{eren2017model}. The benefit of MPC comes from the fact that it selects the optimal control inputs via a user-defined optimization cost function with the capability of incorporating state and input control constraints while optimizing over a receding-time-horizon, subject to expected disturbances and state response. The application of MPC to the LEO magnetic actuation control problem has been explored using only magnetic torque rods~\cite{krogstad2005explicit,wood2008regulation,chen2010}, as well as under-actuated systems with two reaction wheels as opposed to three or four~\cite{petersen2017model}. Nonlinear control of dual-spin and bias-momentum spacecraft via magnetic actuation has also been investigated~\cite{de2012magnetic,de2016spacecraft}. Ref.~\cite{basak2024} presented an LTV (linear time-varying) MPC policy for the stabilization of a magnetically actuated dual-spin spacecraft, and applied their policy to an LTV simulation that includes a gravity gradient torque. 

This paper expands on the literature by introducing complex constraints to the dual-spin stabilized system, including minimum and maximum roll rate limitations, second-order cone constraints on the spacecraft's pointing vector, and the inclusion of soft constraints that allow for continued feasibility in the MPC problem. Crucially, this expansion enables implementation of dual-spin satellites beyond a regulation problem to allow for state deviations that are acceptable within the pointing constraints of the EXACT mission. Further, the predictive control policies are tested in a nonlinear simulation that includes three important disturbance torques: gravity gradient, aerodynamic, and residual magnetic dipole torques. The iterative approach to the LTV predictions used in the MPC policies in this paper provides improved prediction accuracy that likewise improves performance, while maintaining computational tractability through a formulation as a second-order cone program (SOCP). This is especially important for dual-spin satellites, as the magnetic field (and thus the actuation available) is continually varying relative to the satellite even over short prediction horizons.  

As an extension on the work presented in Ref.~\cite{halverson2024}, this paper explores the application of MPC with time-varying prediction models to a magnetically actuated dual-spin stabilized CubeSat in LEO. Ref.~\cite{halverson2024} completed an initial investigation into applying an LTI predictive control policy to this system and compared the performance to a LQR control law. LTV MPC is a well-studied method for control of relevant robotic and other autonomous systems, such as self-driving vehicles~\cite{lima2017} and robotic hexapod manipulators~\cite{qazani2019}. The objectives of this work center around ensuring that the CubeSat's pointing vector is constrained within some allowable drift cone such that the science instrument is continually capable of collecting useful data, and that the system maintains a positive spin rate to maintain a dual-spin configuration. The satellite's control system is allowed (and encouraged) to maximize the drift within prescribed pointing constraints such that the torque required of the magnetic torque rods is minimized. 

To the best of the knowledge of the authors, this paper presents the first predictive control policy for magnetically actuated dual-spin stabilized satellites that (1) enables reliable operation in the presence of real-world nonlinearities, while remaining computationally tractable through a successively linearized time-varying prediction model; (2) is validated in simulation over the entire expected science mission duration of hundreds of orbits for a CubeSat; and (3) allows for drift within specified pointing requirements to meet relevant mission requirements while limiting actuation usage, and thus, the power requirements from the ADCS system. The proposed MPC policies herein are capable of achieving this using a reasonable amount of computational resources to make them suitable for application in small satellite and other space vehicle missions.

Following the preliminaries, this paper proceeds with a discussion of the attitude dynamics of a 3U dual-spin stabilized CubeSat. The environmental model is introduced that is used to describe the disturbance torques a satellite in LEO may encounter, such that an accurate simulation environment can be used for assessment of the control policies. The predictive control policies are introduced, including four MPC policies with different methods of propagation. These policies are compared and results are introduced for several two-orbit simulations in order to quantify the control policies under several different initial conditions. Finally, long simulations are completed to demonstrate the capabilities of the LTV MPC policy to maintain feasibility within the constraint set subject to disturbances in LEO. 

\section{Preliminaries}
This section presents the notation used throughout this paper, as well as a discussion on the important reference frames used to define the problem. 

\subsection{Notation}
The identity matrix is defined by $\mbf{1}\in\mathbb{R}^{n\times n}$, meanwhile the notation $\mbf{1}_m\in\mathbb{R}^{n\times1}$ defines a column vector of zeros with the value in the $m^\text{th}$ position equal to 1, e.g., $\mbf{1}_1 = \left[1 \,\, 0 \,\, 0\right]^\trans$. The cross operator, $(\cdot)^\times: \mathbb{R}^3 \to \mathfrak{so}(3)$, is defined as
\bdis
\mbf{a}^\times = -\mbf{a}^{\times^\trans} = \bbm 0 & -a_3 & a_2 \\ a_3 & 0 & -a_1 \\ -a_2 & a_1 & 0 \ebm,
\edis
where $\mbf{a}^\trans = [ a_1 \,\,\, a_2 \,\,\, a_3 ]$ and $\mathfrak{so}(3) = \{\mbf{S} \in \mathbb{R}^{3 \times 3} \, | \, \mbf{S} + \mbf{S}^\trans = \mbf{0}\}$. 

$\mathcal{F}_a$ is a reference frame defined by a set of three orthonormal dextral basis vectors, $\underrightarrow{a}^1$, $\underrightarrow{a}^2$, and $\underrightarrow{a}^3$. The angular velocity of $\mathcal{F}_b$ relative to $\mathcal{F}_a$ is $\underrightarrow{\omega}^{ba}$. The physical vector describing the position of $p$ relative to $q$ is $\underrightarrow{r}^{pq}$. The time derivative of a physical vector taken with respect to $\mathcal{F}_a$ is expressed as $(\cdot)^{\cdot a}$. The mapping between a physical vector resolved in different reference frames is given by a direction cosine matrix (DCM) $\mbf{C}_{ba}$, where $\mbf{C}_{ba} \in \mathbb{R}^{3x3}$, $\mbf{C}_{ba}\mbf{C}_{ba}^\trans = \mbf{1}$, and det($\mbf{C}_{ba}$) = +1. For example, $\mbf{u}_b = \mbf{C}_{ba}\mbf{u}_a$, where $\mbf{u}_a$ is $\underrightarrow{u}$ expressed in $\mathcal{F}_a$, $\mbf{u}_b$ is $\underrightarrow{u}$ expressed in $\mathcal{F}_b$, and $\mbf{C}_{ba}$ represents the attitude of $\mathcal{F}_b$ relative to $\mathcal{F}_a$. The DCM $\mbf{C}_{i}(\theta)$ is a principle rotation around the $i^{\textrm{th}}$ basis vector by the angle $\theta$.

\subsection{Important Reference Frames}
The Earth-Centered-Inertial (ECI) frame is defined as a non-rotating frame, $\mathcal{F}_a$, where $\ura{a}^1$ points outwards from zero latitude and longitude at J2000, $\ura{a}^3$ points through the north pole, and $\ura{a}^2$ completes the dextral reference frame. The spacecraft body frame, $\mathcal{F}_b$, is defined where $\ura{b}^1$ points through the satellite boresight axis---the long axis of the CubeSat. Basis vectors $\ura{b}^2$ and $\ura{b}^3$ complete the orthonormal dextral reference frame. For the purposes of visualization and linearization, the DCM describing the rotation between $\mathcal{F}_a$ and $\mathcal{F}_b$ is given by the 1-2-3 Euler angle sequence $\mbf{C}_{ba} = \mbf{C}_3(\theta_3)\mbf{C}_2(\theta_2)\mbf{C}_1(\theta_1)$. Within the nonlinear simulation, a quaternion parameterization is used. 

\section{Dual-Spin Configuration Attitude Dynamics \& Kinematics}\label{sec:dynamics}
The equations of motion describing the attitude dynamics of a dual-spin stabilized spacecraft are
\begin{equation}\label{eq:DSS_full}
    \mbf{I}_b^{Bc}\mbs{\omega}_b^{ba^{\cdot a}} + \mbs{\omega}_b^{ba^{\times}}\left(\mbf{I}_b^{Bc}\mbs{\omega}_b^{ba} + \mbf{a}_s h_s\right) + \mbf{a}_s \dot{h}_s = \mbs{\tau}_b^{\text{mag}} + \mbs{\tau}_b^{\text{ext}},
\end{equation}
where $\mbf{I}_b^{Bc}$ is the moment of inertia of the spacecraft resolved in $\mathcal{F}_b$; $\mbs{\omega}_b^{ba}$ is the angular velocity of the spacecraft relative to $\mathcal{F}_a$ resolved in $\mathcal{F}_b$; $h_s = I_s \omega_s$ is the wheel spin angular momentum relative to the spacecraft with $I_s$ and $\omega_s$ describing the inertia and angular velocity of the momentum wheel, respectively; $\mbf{a}_s$ denotes the momentum wheel spin axis; $\mbs{\tau}_b^\text{mag}$ is the torque caused by actuation of the magnetic torque rods; and $\mbs{\tau}_b^\text{ext}$ includes all external disturbance torques acting on the spacecraft. A thorough description of these equations of motion, stability of dual-spin satellites in torque-free motion, and a discussion on the effect of internal energy dissipation can be found in Ref.~\cite{DeRuiter2013}. 

\begin{figure}[t!]
	\includegraphics[width=0.45\linewidth]{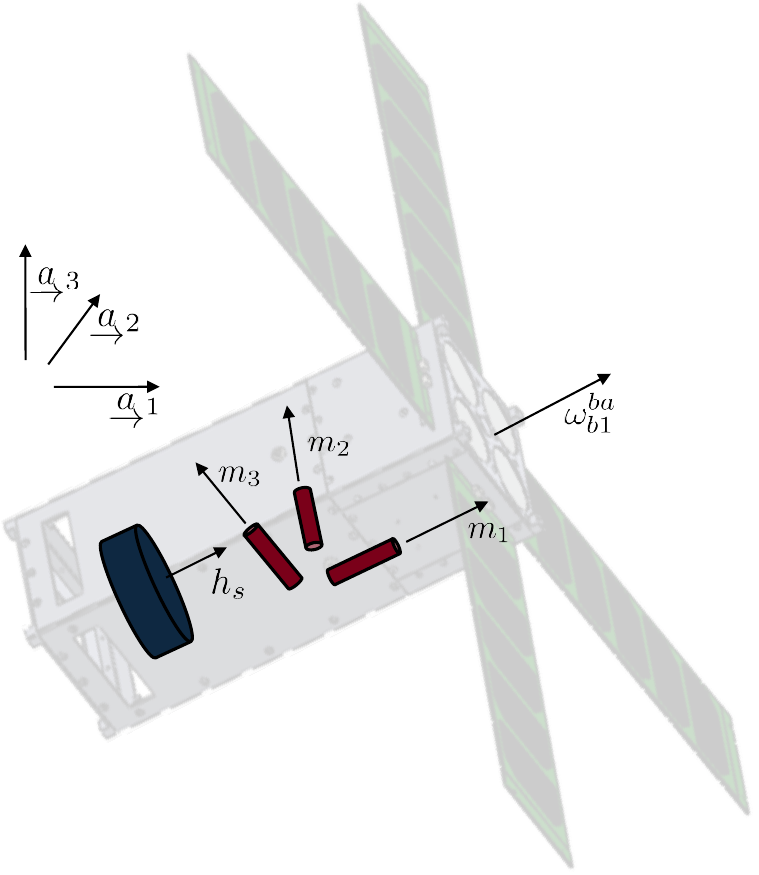} 
	\centering
	\caption{Dual-spin stabilized CubeSat actuator configuration. }\label{fig:DSS_diagram}
\end{figure}

A diagram depicting the dual-spin configuration with magnetic torque rods is included in Fig.~\ref{fig:DSS_diagram}, where the CubeSat is transparent to show the control hardware within. The reaction wheel is depicted in blue with momentum $h_s$ and the angular velocity, $\omega_{b1}^{ba}$, is aligned with the science instrument boresight axis. The torque rods are aligned in the orthonormal directions of the basis vectors that define reference frame $\mathcal{F}_b$, i.e., the direction of the magnetic dipoles of the torque rods are aligned with the basis vectors $\ura{b}^i$. The nominal angular velocity for the dual-spin stabilized configuration is defined as $\mbs{\omega}_{b,\text{nom}}^{ba} = \left[\gamma~0~0\right]^\trans$, where $\gamma$ is the nominal spin rate, and the momentum wheel spin axis is $\mbf{a}_s = \left[1~0~0\right]^\trans$. It is important to note that in this paper, a variable-speed momentum wheel is utilized and as such $\dot{h}_s$ is an additional control input.

For the nonlinear simulation, a quaternion parameterization is used to describe the satellite's attitude. This results in the attitude kinematics being defined as
\begin{equation*}
    \dot{\mbf{q}} = \bbm\dot{\mbs{\epsilon}} \\ \dot{\eta} \ebm = \frac{1}{2} \bbm \eta\mbf{1}+\mbs{\epsilon}^\times \\ -\mbs{\epsilon}^\trans \ebm\mbs{\omega}_b^{ba}.
\end{equation*}
A complete discussion of this attitude parameterization is omitted for brevity. However, a thorough discussion of the quaternion and its suitability in space vehicle applications can be found in Refs.~\cite{de2016spacecraft,hughes2012spacecraft}

\subsection{Magnetic Actuation Dynamics} \label{sec:dyn_mag}
The actuator dynamics that describe the torque rod control input in Eq.~\eqref{eq:DSS_full} are defined as
\begin{equation}\label{eq:tr_dynamics}
    \mbs{\tau}_b^\text{mag} = \mbf{m}_u^\times \mbf{b}_b,
\end{equation}
where $\mbf{m}_u = \left[m_1 \,\,\, m_2 \,\,\, m_3 \right]^\trans$ is the torque rod control input, and $\mbf{b}_b$ is the Earth's magnetic flux density---the direction and magnitude of the magnetic flux from the Earth's magnetic field. Magnetic torque rods are coiled material (usually copper wire) that are actuated through an input of current through the rod. The magnetic dipole can be expressed as $m_i = nI_iA$, where $n$ is the number of turns of the wire, $I_i$ is the current, and $A$ is the effective vector area. 

Many of the challenges associated with magnetic actuation are due to the reliance on the orientation of the satellite relative to Earth's magnetic field and the limited capability of actuation. The Earth's magnetic field is very weak, and the magnitude of torque the rods can generate is very small---thus providing limited control authority. Further, this actuation method alone is incapable of inducing a torque around the direction of the magnetic field vector. This is observed in Eq.~\eqref{eq:tr_dynamics}, such that the cross-product limits the torque capable of being generated to be perpendicular to the instantaneous magnetic field vector. In non-equatorial orbits, 3-axis controllability is obtainable over the course of the entire orbit due to the direction of Earth's magnetic field line varying relative to the spacecraft such that the satellite is able to produce a torque in all axes.

\subsection{Low-Earth Orbit Attitude Disturbance Model}\label{sec:dyn_dist}
There are several disturbance torques that satellites in LEO experience that must be taken into account when developing a realistic simulation and effective control policy. This work incorporates the most prevalent disturbances including magnetic, gravitational, and aerodynamic torques. These disturbances, respectively, define $\mbs{\tau}_b^{\text{ext}}$ in Eq.~\eqref{eq:DSS_full};
\begin{equation*}
    \mbs{\tau}_b^\text{ext} = \mbs{\tau}_b^\text{magd} + \mbs{\tau}_b^\text{grav} + \mbs{\tau}_b^\text{aero} = \mbf{m}^{\text{d}^\times}_b \mbf{b}_b + \frac{3\mu}{||\mbf{r}_b^{ce}||^5} \mbf{r}_b^{ce^\times}\mbf{I}_b^{Bc}\mbf{r}_b^{ce} + \mbf{c}^{\text{p}^\times}_b \mbf{F}_b^\text{aero},
\end{equation*}
where $\mbf{m}^\text{d}_b$ is the residual magnetic dipole vector of the spacecraft, $\mu$ is the gravitational parameter of Earth, $\mbf{c}^\text{p}_b$ is the location of the spacecraft's center of pressure relative to its center of mass, and $\mbf{F}_b^\text{aero}$ is the aerodynamic drag force defined by the projected area of the satellite surface in the direction of velocity. A great amount of detail regarding these and other much less significant disturbances can be found in Ref.~\cite{hughes2012spacecraft}, along with a discussion on these disturbances specific to this problem in Ref.~\cite{halverson2024}.

\subsection{Uncompensated Pointing Error}\label{sec:results_OL}
To demonstrate the nonlinear dynamical simulation with the disturbances outlined in this section, a hypothetical scenario is implemented in two simulations. A two-orbit simulation begins with significant attitude and angular rate errors with an initial roll-pitch-yaw attitude of $\mbs{\Theta}^\trans = \left[ \theta_1 \,\, \theta_2 \,\, \theta_3 \right] = \left[0 \,\, 4.5 \,\, -6.5 \right]$ degrees, and an initial angular velocity of $\mbs{\omega}_b^{ba\trans} = \left[ \gamma \,\, 0.3 \,\, -0.25 \right]$ degrees/second, where $\gamma = 0.75$~degrees/second is the nominal roll rate. The satellite used in this simulation is symmetric about the $b_1$ direction with a reaction wheel spin rate of $400$~rad/s. Other details regarding the specific simulation qualities are outlined in Section~\ref{sec:results}. 

\begin{figure}[t!]
	\centering
	\begin{subfigure}[]{0.5\textwidth}
		\includegraphics[scale=0.63]{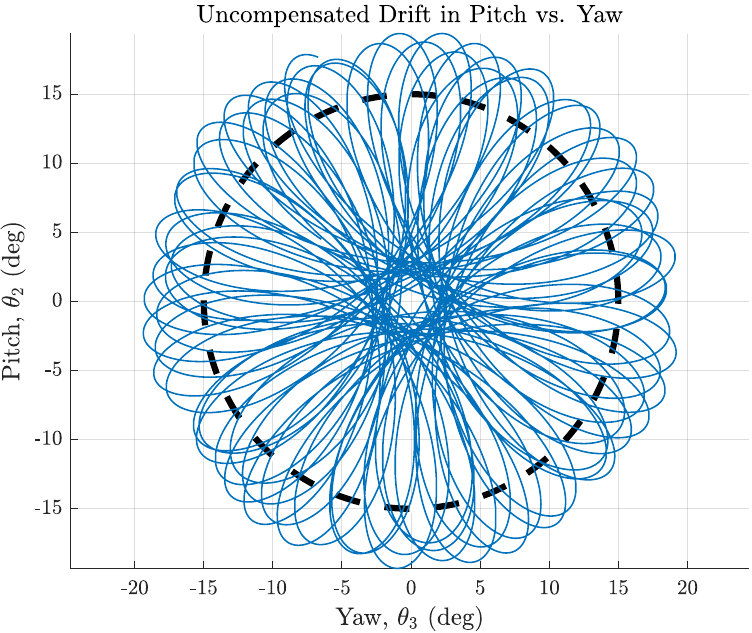}
        \caption{2-orbit simulation}\label{fig:OL_sims2orb}
        \centering
	\end{subfigure}%
	\begin{subfigure}[]{0.5\textwidth}
		\includegraphics[scale=0.63]{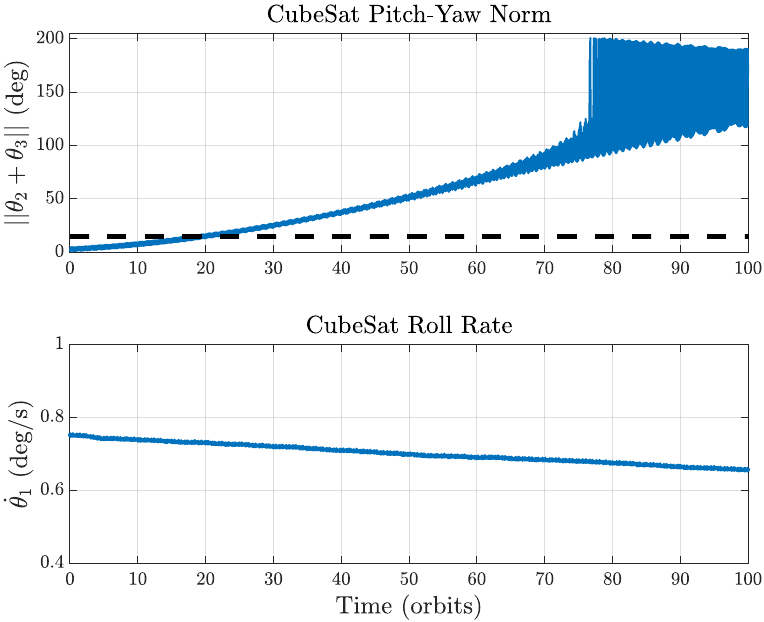} 
        \caption{100-orbit simulation}\label{fig:OL_sims100orb}
        \centering
	\end{subfigure}%
	\centering
	\vspace{-8pt}
	\caption{Uncompensated state drift for open-loop simulations.}\label{fig:OL_sims}
\end{figure}

Results from the two-orbit simulation are shown in Fig.~\ref{fig:OL_sims2orb}. The initial conditions provide a response such that the uncontrolled spacecraft states exceed constraint bounds quickly, most prevalently the pointing cone constraint with a maximum allowable half-cone angle of $15^\circ$ shown in a black dashed line. These initial conditions are realistic to imperfect conditions and residual angular velocity from a combination of tip-off rates and disturbance torques.

A simulation of 100 orbits with initial conditions closer to the nominal configuration is also completed, with results on the pointing norm of pitch and yaw angles along with the roll rate shown in Fig.~\ref{fig:OL_sims100orb}. The initial conditions used are an initial roll-pitch-yaw attitude of $\mbs{\Theta}^\trans = \left[0 \,\, 1 \,\, -0.5 \right]$ degrees, and an initial angular velocity of $\mbs{\omega}_b^{ba\trans} = \left[ \gamma \,\, 0.01 \,\, -0.075 \right]$ degrees/second. This data demonstrates that even with initial conditions that are more favorable for the half-cone angle constraint, the satellite still violates this around 20 orbits. Meanwhile, the roll rate also decreases slowly, likely due to the dissipation of energy due to external disturbances. 

The useful characteristics of a dual-spin stabilized configuration are clearly presented in these results, as the satellite's pointing vector oscillates back and forth, swept out in a circular motion due to the nominal spin of the satellite. This is a product of the passive stability properties and gyroscopic stiffness of dual-spin stabilization. Section~\ref{sec:MPC} aims to bound this response within the specified constraints while minimizing control input, where different MPC policies are introduced and compared. 

\subsection{Linearization about a Nominal Spin}\label{sec:dyn_lin}
Given the simplicity of the inertial pointing mission, an Euler angle attitude parameterization is used for the linearization required for the MPC policies. The kinematic expression to relate Euler angles to angular velocity is defined as $\mbs{\omega}_b^{ba} = \mbf{S}_b^{ba}(\mbs{\Theta})\dot{\mbs{\Theta}}$, where
\begin{equation}\label{eq:S_mat}
    \mbf{S}_b^{ba}\left(\mbs{\Theta}\right) = \bbm
    \cos(\theta_3)\cos(\theta_2) & \sin(\theta_3) & 0 \\
    -\sin(\theta_3)\cos(\theta_2) & \cos(\theta_3) & 0 \\
    \sin(\theta_2) & 0 & 1\ebm
\end{equation}
is the mapping matrix between angular rates and angular velocity. This is chosen as a 1-2-3 Euler angle rotation sequence, which not only places the kinematic singularity at $180^\circ$ from the nominal inertial pointing attitude, but importantly allows for a simple linearization about a nominal roll rate due to $\mbf{S}_b^{ba}\left(\mbs{\Theta}\right)$ not being a function of $\theta_1$. Assuming a generic trajectory of Euler angles ($\mbs{\Theta}(t)$) is available, the time-varying kinematic mapping from Euler angle rates to angular velocity exists simply as $\mbs{\omega}_b^{ba} = \mbf{S}_b^{ba}\left(\theta_{2}(t),\theta_{3}(t)\right)\dot{\mbs{\Theta}}$.

Two key assumptions are made in the linearization of the satellite dynamics. First, it is assumed that the spacecraft body axes ($\mathcal{F}_b$) are aligned with the principal axes of inertia such that $\mbf{I}_b^{Bc} = \text{diag}\{I_1,I_2,I_3\}$. Second, the nominal spin rate of the satellite, $\gamma$, is assumed to be constant, and thus $\dot{\gamma} = 0$. The perturbations from the equilibrium angular velocity, $\mbs{\omega}_{b,\text{nom}}^{ba}$, is denoted by $\delta\mbs{\omega}_b^{ba} = \left[(\gamma + \delta\omega_1) \,\,\, \delta\omega_2 \,\,\, \delta\omega_3 \right]^\trans$.
Replacing the angular velocity in Eq.~\eqref{eq:DSS_full} with these perturbed states, removing higher-order perturbation terms, and setting $\dot{\gamma}$ to zero, three scalar equations are recovered such that
\begin{align*}
I_1\delta\dot{\omega}_1 + I_s\dot{\omega}_s &= 0, \\
I_2\delta\dot{\omega}_2 + (I_1-I_3)\gamma\delta\omega_3 + I_s\omega_s\delta\omega_3 &= 0, \\
I_3\delta\dot{\omega}_3 + (I_2-I_1)\gamma\delta\omega_2 + I_s\omega_s\delta\omega_2 &= 0,
\end{align*}
which completes the linearization of the attitude dynamics for a nominal spin around the $\ura{b}^1$ axis.

\subsubsection{State-Space Realization of LTV Equations of Motion}\label{sec:dyn_stateSpace}
The system states are defined as $\mbf{x} = \left[ \delta\mbs{\Theta}^\trans \,\,\, \delta\mbs{\omega}_b^{ba\trans} \right]^\trans$ and control inputs as $\mbf{u} = \left[ \dot{\omega}_s \,\,\, m_1 \,\,\, m_2 \,\,\, m_3 \right]^\trans$. The linearized model is written in an LTV state space representation as
\begin{equation}\label{eq:CT_LTV}
    \dot{\mbf{x}} = \mbf{A}(t)\mbf{x}(t) + \mbf{B}_u(t)\mbf{u}(t) + \mbf{B}_w\mbf{w}(t),
\end{equation}
where
\beq
    \mbf{A}(t) = \bbm
            -\gamma(t)\mbf{1}_1^\times & \mbf{S}_b^{ba}\left(\theta_{2}^{(i)}(t),\theta_{3}^{(i)}(t)\right) \\
            \mbf{0}_{3\times3} & \mbf{A}_\text{dss}(t) \ebm, \nonumber 
    \hspace{8pt}
    \mbf{B}_u = \bbm
            \mbf{0}_{3\times1} & \mbf{0}_{3\times3} \\
            \frac{-I_s}{I_1}\mbf{1}_1 & \mbf{b}_b^\times(t) \ebm, \nonumber
    \hspace{8pt}
   \mbf{B}_w = \bbm 
            \mbf{0}_{3\times3} \\
            \mbf{1}_{3\times3}\ebm,\nonumber
    \hspace{8pt}
    \mbf{A}_\text{dss}(t) = \bbm
			0 & 0 & 0 \\
			0 & 0 & \frac{\sigma_1(t)}{I_2} \\
			0 & \frac{\sigma_2(t)}{I_3} & 0 \ebm,\nonumber
\eeq
$\mbf{w}(t) = \tau_b^\text{ext}(t)$ includes all disturbance torques, $\sigma_1(t) = (I_3-I_1)\gamma(t)-I_s\omega_s(t)$, and $\sigma_2(t) = (I_1-I_2)\gamma(t)+I_s\omega_s(t)$.

The continuous-time LTV state-space system in Eq.~\eqref{eq:CT_LTV} is converted to a discrete-time LTV state-space realization with time step $\Delta t$ of the form
\begin{equation*}
	\mbf{x}_{k+1} = \mbf{A}_{\text{d},k} \mbf{x}_k + \mbf{B}_{u,\text{d},k} \mbf{u}_k + \mbf{B}_{w,\text{d}}\mbf{w}_k, 
\end{equation*}
where the subscript $k$ denotes the index of the discrete time step. A zero-order hold is used to perform the discretization.

\subsubsection{Controllability Analysis}
An LTI system is said to be controllable when at any time, $t_0$, it is possible to find some sequence of inputs that will transfer any initial state, $\mbf{x}(t_0)$, to the origin in finite time~\cite{brogan1991modern}. Controllability of these systems can be proven when the rank of the controllability matrix, 
\begin{equation*}
    \mathcal{C} = \bbm \mbf{B} & \mbf{A}\mbf{B} & \mbf{A}^2\mbf{B} & \dots & \mbf{A}^{n-1}\mbf{B} \ebm,
\end{equation*}
has rank $n$ and $\mbf{A}\in \mathbb{R}^n$. In other words, the matrix $\mathcal{C}$ must have $n$ linearly independent columns such that it spans all of $\mathbb{R}^n$. In the case of an LTV system, there are rigorous methods to determine the controllability of which some make use of integration of the state transition matrix over the expected time-varying parameters. The reader is encouraged to read Ref.~\cite{silverman1967} for details on theoretical proofs of controllability and observability of LTV systems. 

A thorough theoretical controllability analysis may not be necessary when leveraging physical understanding of the problem, where instead dynamic intuition is used to determine the worst-case conditions and perform an LTI controllability analysis. Specifically, we select conditions where magnetic actuation alone does not have instantaneous controllability and update them with the dynamics corresponding with the dual-spin stabilized configuration. In each case where LTI dynamics are used for this analysis, we assume $\mbs{\Theta}(t) = \mbf{0}_{3\times1}$, $\dot{\mbf{b}}_b(t) = \mbf{0}_{3\times1}$, $\dot{\gamma}(t) = 0$, and $\dot{\omega}_s(t) = 0$. 

Using only magnetic actuation with no nominal spin (i.e., $\gamma = 0$), $\mathcal{C}$ is rank deficient in all orientations of the spacecraft. Physical intuition explains that no torque is available about the direction of the magnetic field vector. When a constant non-zero spin rate is implemented, complete instantaneous controllability is achieved in all satellite configurations but one. This is shown analytically using the LTV dynamics in Eq.~\eqref{eq:CT_LTV} at any specific instance in time, which are LTI based on the assumptions above, where $\text{Rank}(\mathcal{C})<n$ when $\gamma = 0$, and $\text{Rank}(\mathcal{C})=n$ when $\gamma \neq 0$. 

In contrast to the representative mission described in this paper of a dual-spin satellite equipped with a variable-speed reaction wheel, a nominal dual-spin stabilized configuration often uses a constant angular-rate momentum wheel. This uncovers a singular worst-case orientation when the spin axis aligns with the Earth's magnetic field vector at a specific time. At this attitude and time, actuation is instantaneously unavailable about the direction of the satellite's spin axis (the same direction as the momentum wheel's spin axis). By equipping the satellite with a variable-speed reaction wheel, the system recovers full instantaneous controllability in this worst-case scenario, proven both via empirical understanding and analytically through $\text{Rank}(\mathcal{C})=n$ when $\gamma \neq 0$ and the capacity for $\dot{\omega}_s \neq 0$. Thus, the magnetically actuated LTV system is guaranteed to have continual instantaneous controllability throughout the lifetime of the satellite so long as a nominal dual-spin stabilized configuration is maintained with a variable-speed reaction wheel. 

\section{Model Predictive Control Framework}\label{sec:MPC}
MPC is a suitable control approach for when system and environment dynamics are well understood and when there exist specific constraints that not only need to be met, but can also be utilized to the advantage of the controller. In the problem presented in this paper, the inertial pointing mission is operating in the familiar LEO environment with simple and convex constraints that, among others, include a region of allowable attitude deviation, making MPC a suitable control approach. 

Three novel LTV MPC policies are described in this section, including a policy that incorporates only the changing magnetic field due to orbital position, an iterative policy that propagates the LTV dynamics using the control solution of the previous iteration to capture the time-varying attitude dynamics, and a nonlinear-propagated iterative policy that further enhances the prediction quality. These LTV policies are expected to have a tradeoff between computational efficiency and the required magnetic actuation along with constraint satisfaction. Thus, we investigate this hypothesis via a comparison of the prediction accuracy and later challenge them in a nonlinear simulation of the system and environment. The LTV policies are further compared to an LTI MPC policy that completely foregoes all time-varying aspects of the problem and instead uses a constant magnetic field vector prediction similar to the policy presented in Ref.~\cite{halverson2024}, which aims to demonstrate the necessity of including the LTV dynamics in the MPC prediction.

Each MPC policy uses the same base receding-horizon optimal control problem with a quadratic cost and linear and quadratic constraints. The general MPC optimization problem is solved at each timestep in the prediction horizon with length $\Delta t$, and is defined as
\begin{equation}\label{eq:Q2-MPC}
\min_{\mathcal{U}_t,\mathcal{X}_t,\mathcal{V}_i} \hspace{5pt} \sum_{k=0}^{N-1}\left(\mbf{x}_{k|t}^\trans\mbf{Q}\mbf{x}_{k|t} + \mbf{u}_{k|t}^\trans\mbf{R}\mbf{u}_{k|t} + \mbs{\Psi}_{\nu_1}\mbs{\nu}_1 + \mbs{\Psi}_{\nu_2}\mbs{\nu}_2 + \mbs{\Psi}_{\nu_3}\mbs{\nu}_3\right),
\end{equation}
subject to
\begin{align}
&\mbf{x}_{k+1|t} = \mbf{A}_{\text{d},k}\mbf{x}_{k|t} + \mbf{B}_{u,\text{d},k}\mbf{u}_{k|t} + \mbf{B}_{w,\text{d}}\hat{\mbf{w}}_{k|t}, \label{eq:MPC_prediction} \\
&\mbf{x}_{0|t} = \mbf{x}(t), \nonumber \\
&\omega^{ba}_{b1,k|t} \geq \gamma_{\text{min}}, \nonumber \\
&\mbf{g}_i(\mbf{x}_{k|t}) \leq \mbs{\nu}_i, \quad \mbs{\nu}_i \geq \mbf{0}, \nonumber \\
&\mbf{u}_{\text{min}} \leq \mbf{u}_{k|t} \leq \mbf{u}_{\text{max}}, \quad 0 \leq k \leq N-1,\label{eq:MPC_cConstraint} 
\end{align}
where $N$ is the prediction horizon length, $\mathcal{U}_t = \{\mbf{u}_{0|t},\ldots,\mbf{u}_{N-1|t}\}\in \mathbb{R}^{4\times N}$, $\mathcal{X}_t = \{\mbf{x}_{0|t},\ldots,\mbf{x}_{N|t}\}\in \mathbb{R}^{6\times (N+1)}$, $\mathcal{V}_i = \{\mbs{\nu}_{1},\mbs{\nu}_{2},\mbs{\nu}_{3}\}\in \mathbb{R}^{N\times 3}$, $\mbf{Q} = \mbf{Q}^\trans \geq 0$ and $\mbf{R} = \mbf{R}^\trans > 0$ are constant state and control weighting matrices, and $\mbfhat{w}_{k|t} = \mbfhat{w}_t(t+k\Delta t)$ is the open-loop predicted disturbance column matrix from time $t$ to time $N\Delta t$ based on data at time $t$. Once the convex optimization problem in Eq.~\eqref{eq:Q2-MPC} is solved, the optimal control input for the first timestep, denoted as $\mbf{u}_t^*$, is applied to the system in a zeroth-order hold fashion and the MPC problem is solved again at the next timestep given updated state information. The subscript notation $k|t$ denotes the state or control input $k$ prediction steps ahead of time $t$. For example, $\mbf{x}_{k|t}$ is the predicted state $k$ steps ahead of time $t$. 

In each MPC policy, there are several important constraints that must be satisfied for both control saturation and state allowance purposes. The control input constraint in Eq.~\eqref{eq:MPC_cConstraint} includes limitations in control authority for the magnetic torque rods and the reaction wheel. The torque rods are constrained to satisfy $\mbf{u}_{m,\text{min}} \leq \mbf{m}_u \leq \mbf{u}_{m,\text{max}}$, where $\mbf{u}_{m,\text{min}} = -\mbf{u}_{m,\text{max}}$ are the torque rod saturation limits. It is assumed the torque rods are able to supply a continuously variable input. The constraint on the reaction wheel input torque is defined as $u_{w,\text{min}} \leq \dot{\omega}_s \leq u_{w,\text{max}}$, where $u_{w,\text{min}} = -u_{w,\text{max}}$ is the largest change in angular velocity of the reaction wheel. 

There are two sets of constraints acting on the roll rate ($\omega^{ba}_{b1,k|t} = \dot{\theta}_{1,k|t}$) of the satellite. A hard constraint, $\omega^{ba}_{b1,k|t} > \gamma_\text{min}$, ensures the roll rate does not cross zero to maintain spin stability, nor does it swap to a rate opposite of the linearization polarity. Meanwhile, two soft constraints are also employed, where the first limits the maximum to be of a reasonable magnitude (within the trust region of the linearization), while the second acts as a barrier to the hard constraint by limiting the minimum roll rate to be reasonably nonzero and protect the system from falling into a region of infeasibility in the control optimization problem. Specifically, these soft constraints are defined as
\begin{align*}
    \mbf{g}_1(\mbf{x}_{k|t}) &= \omega^{ba}_{b1,k|t} - \gamma_\text{max,s}, \\
    \mbf{g}_2(\mbf{x}_{k|t}) &= \gamma_\text{min,s} - \omega^{ba}_{b1,k|t},
\end{align*}
where $\gamma_\text{max,s} = -\gamma_\text{min,s}$ are the maximum and minimum roll rate soft constraints, respectively.

In order to utilize the entire drift allowance in the pointing cone constraint (illustrated in Fig.~\ref{fig:DSS_pointing}) and limit the possibility for the optimization problem falling within an area of infeasibility, a cone soft constraint is implemented. This constraint is defined as 
\begin{equation}\label{eq:cone_constraint}
    \mbf{g}_3(\mbf{x}_{k|t}) = \norm{\left[ \theta_{2,k|t} \,\,\, \theta_{3,k|t} \right]}_2 - \alpha_\text{max,s},
\end{equation}
where $\alpha_\text{max,s}$ is the maximum allowable drift of the satellite pointing vector norm.

Each soft constraint includes weights in the cost function that multiply the slack variables, $\nu_i$, to penalize constraint violation. These are denoted by the constant weights $\mbs{\Psi}_{\nu_{i}} = \psi_i \cdot \left[1 \ldots 1\right] \in \mathbb{R}^{N+1}$ for each of the soft constraints, where each cost weight $\psi_i$ is selected according to design requirements in performance and determination of the relative importance of each constraint with one another. The slack variables, $\nu_i$, are nominally zero when the soft constraints are inactive according to the optimal solution, and are otherwise constrained to be positive. 

This MPC policy is in the form of a quadratically-constrained quadratic program (QCQP), which can be formulated as a second-order cone program (SOCP). Techniques to do so can be found in Ref.~\cite{martins2021engineering}. The remainder of this section presents the four different prediction models considered for use with this MPC policy. Open-loop simulation results using each model over an example prediction horizon are presented to demonstrate their effectiveness in capturing the spacecraft's time-varying and nonlinear response. 

\subsection{Constant Magnetic Field Prediction}\label{sec:MPC_const}

The simplest form of a prediction model for this system relies on the assumption that the magnetic field vector remains constant in $\mathcal{F}_b$ over the course of the entire prediction horizon. This is the primary assumption made in the MPC policy implemented in Ref.~\cite{halverson2024}. Specifically, the prediction dynamics in Eq.~\eqref{eq:MPC_prediction} make the assumption that $\mbf{B}_u$ is constant over the entire prediction horizon and hence is evaluated at time $t$. This assumption is not expected to incur much error when the prediction horizon length is short enough. This assumption in no way takes into account the effect that translational motion and attitude variation have on the expected magnetic field direction over time, which then is expected to directly degrade the accuracy of the prediction model's knowledge of the affect magnetic actuation has on the satellite's motion. 

Further, given that no time history of expected attitude states are available, the kinematic mapping from Euler angle rates to angular velocity assumes that small perturbations in angular rates and angular velocity are approximately equivalent. Mathematically, this is described by the expression $\delta\mbs{\omega}_b^{ba} \approx \delta\dot{\mbs{\Theta}}$, where the defining assumption is that $\mbf{S}_b^{ba}(\mbs{\Theta}) = \mbf{1}_{3\times3}$. A complete discussion on this approximation of kinematics is included in Ref.~\cite{halverson2024}.

\begin{figure}[t!]
	\centering
	\begin{subfigure}[]{0.4\textwidth}
		\includegraphics[scale=0.6]{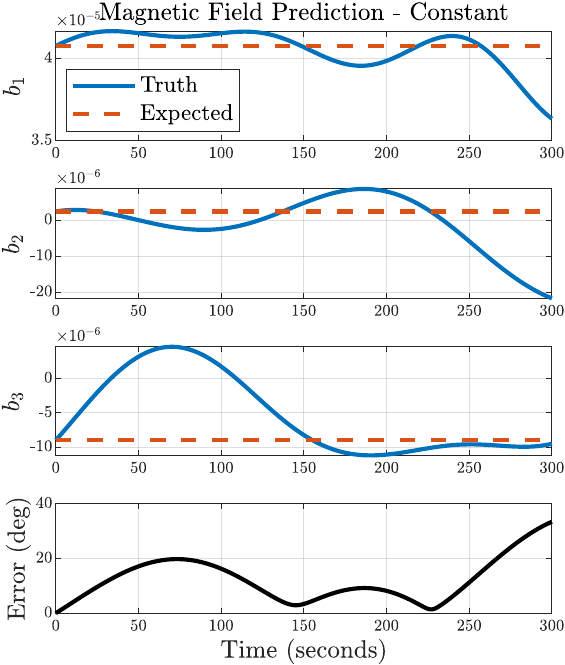}
        \caption{Magnetic field vector components}
        \label{fig:constPredA}
        \centering
	\end{subfigure}%
	\begin{subfigure}[]{0.6\textwidth}
		\includegraphics[scale=0.6]{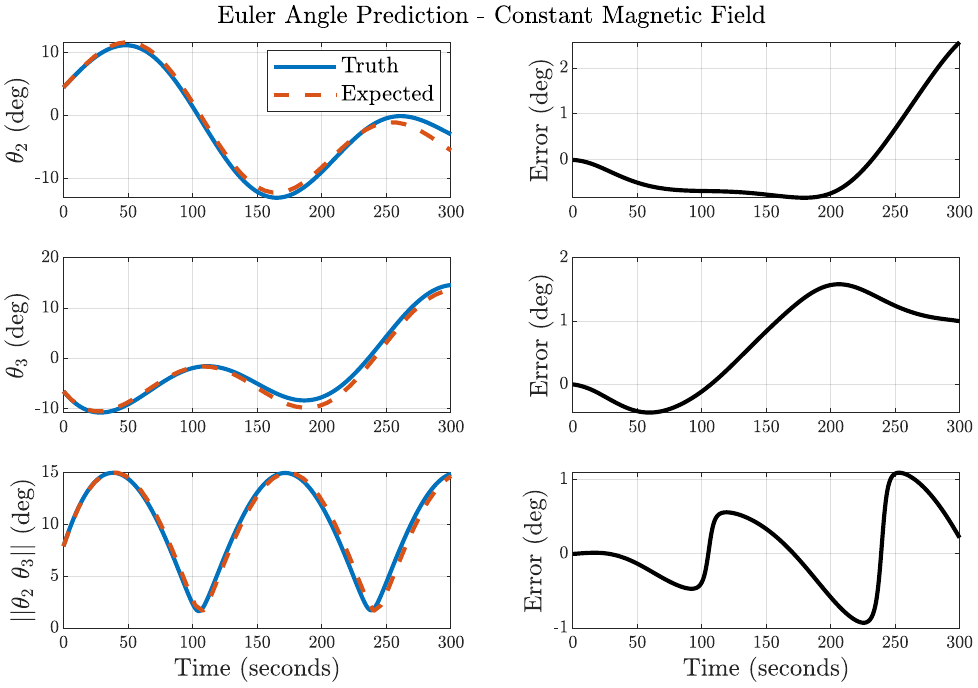}
        \caption{Pitch, yaw, and norm values and errors}
        \label{fig:constPredB}
        \centering
	\end{subfigure}%
	\centering
	\vspace{-8pt}
	\caption{Constant magnetic field prediction method expected values.}\label{fig:Const_pred}
\end{figure}

The magnetic field prediction error and subsequent effect on the expected pitch-yaw pointing are illustrated in Fig.~\ref{fig:Const_pred}, where a sample prediction horizon length of 300 seconds ($N = 50$, $\Delta t = 6$ seconds) is used for the MPC policy in Eq.~\eqref{eq:Q2-MPC} with a constant magnetic field direction assumption. The simulation itself includes the same configuration and initial conditions as described in the uncompensated response illustrated in Fig.~\ref{fig:OL_sims}. Each simulation in the proceeding subsections use the same initial conditions and simulation variables.

In Fig.~\ref{fig:constPredA}, the error is calculated as the angle between the true magnetic field vector that is experienced by the satellite over the simulation and the magnetic field vector that is `expected' by the MPC policy. This calculation is performed via the dot product definition. Each proceeding subsection uses this process for error calculation. Values for pitch, yaw, and the norm of these angles (i.e., the `pointing vector' that is constrained within the allowable half-cone angle) are shown along with the corresponding errors accumulated over the prediction horizon in Fig.~\ref{fig:constPredB}. This result illustrates the effect of a poor prediction quality due to assumptions made in the magnetic field propagation and linear dynamics with an error of near or over $2^\circ$ in both pitch and yaw over the prediction horizon. This error would likely be unacceptable in implementation given that the prediction model cannot be trusted to handle very long horizon lengths, thus further degrading performance of the MPC policy. 

\subsection{Orbital Scheduling Prediction}\label{sec:MPC_orb}

\begin{figure}[t!]
	\centering
	\begin{subfigure}[]{0.4\textwidth}
		\includegraphics[scale=0.6]{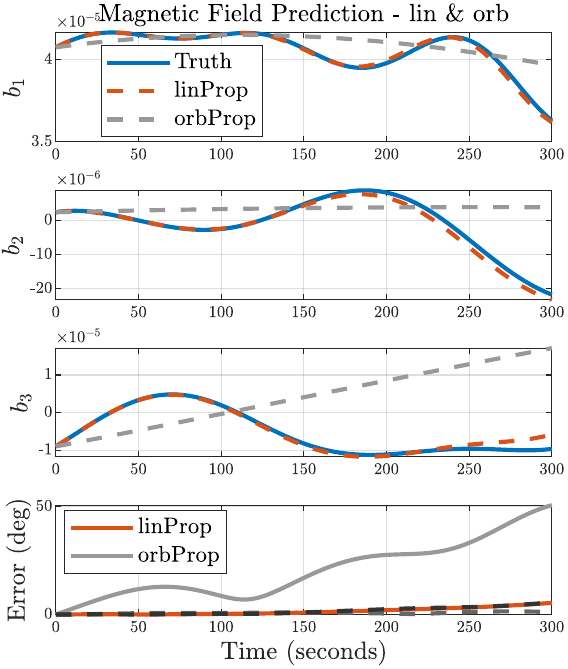}
        \caption{Magnetic field vector components}
        \centering
	\end{subfigure}%
	\begin{subfigure}[]{0.6\textwidth}
		\includegraphics[scale=0.6]{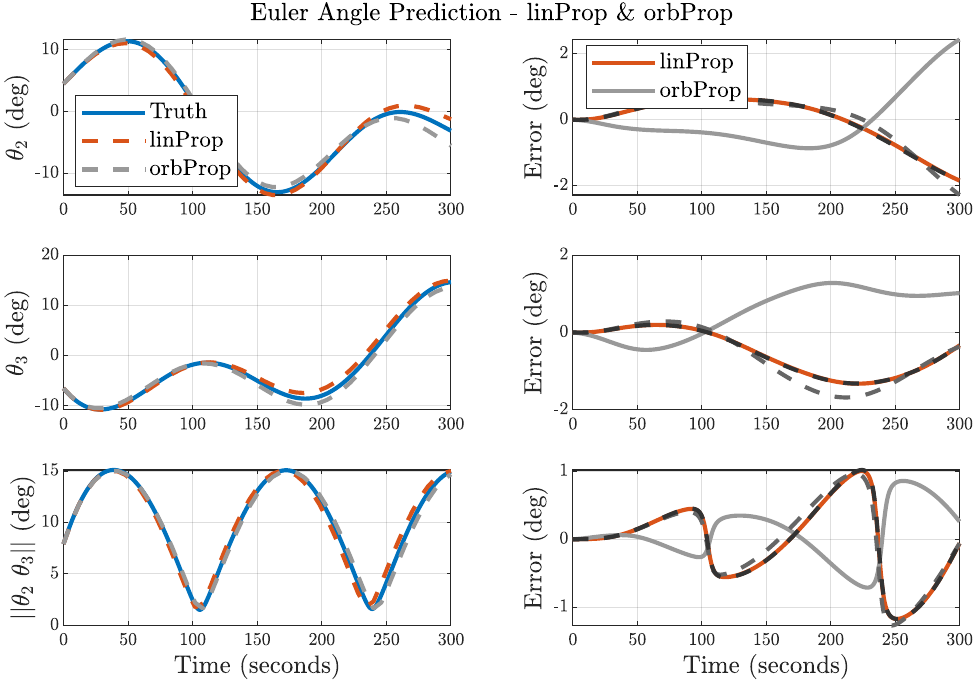} 
        \caption{Pitch, yaw, and norm values and errors}
        \centering
	\end{subfigure}%
	\centering
	\vspace{-8pt}
	\caption{Orbital scheduling (`orbProp') and linear propagation prediction (`linProp') method expected values.}\label{fig:orbLin_pred}
\end{figure}

The prediction scheme in this section takes into account the orbital trajectory of the satellite and its effect on the expected magnetic field vector. Specifically, the translational states at time $t$ are propagated over the prediction horizon according to the undisturbed classical form of Kepler's equation,
\begin{align*}
    M_{k|t} = E_{k|t} - e_{k|t}\sin{E_{k|t}},
\end{align*}
where $M$ is the orbital mean anomaly, $e$ is the eccentricity, and $E$ is the eccentric anomaly. A detailed description of Kepler's equation and how to solve it in terms of position and velocity vectors can be found in Chapter 9.6 in Ref.~\cite{schaubBook}. The attitude of the satellite ($\mathcal{F}_b$) relative to $\mathcal{F}_a$ is assumed to remain constant over the prediction horizon using this prediction model, and the same assumption is used regarding the mapping between angular rates and angular velocity, where $\mbf{S}_b^{ba} = \mbf{1}_{3\times3}$. The resulting magnetic field and attitude errors are illustrated as a portion of the results in Fig.~\ref{fig:orbLin_pred}.

The time-varying nature of the magnetic field vector relative to the satellite solely due to orbital motion is captured by this prediction method, which is expected to add accuracy in the prediction compared to a constant assumption. It is immediately clear that the maximum error in the pitch-yaw norm is slightly smaller than that of the constant assumption case, however the magnetic field prediction error is not improved and the total error accumulated over time is greater. 

\subsection{Linear Propagation Prediction}\label{sec:MPC_lin}

This method takes into account the time-varying aspect of the attitude of the satellite and its effect on the magnetic actuation available. In order to account for the changing attitude, an iterative approach is used. An initial guess to iterate upon is selected as the solution from the orbital scheduling prediction from Section~\ref{sec:MPC_orb}. 

Given a nonzero state trajectory for the satellite's attitude, complete discrete-time LTV equations of motion including all time-varying parameters can be used as defined in Section~\ref{sec:dyn_lin}. Thus, the dynamics used in this MPC prediction formulation are realized as $\mbf{x}_{k+1}^{(i)} = \mbf{A}_{\text{d},k}^{(i)} \mbf{x}_k^{(i)} + \mbf{B}_{u,\text{d},k}^{(i)}\mbf{u}_k^{(i)} + \mbf{B}_{w,\text{d}}\mbf{w}_k$, where the continuous-time dynamics matrix at the $i^{th}$ iteration is specifically defined as
\begin{equation}\label{eq:A_LTV}
    \mbf{A}^{(i)}(t) = \bbm 
        -\gamma^{(i)}(t)\mbf{1}_1^\times & \mbf{S}_b^{ba}\left(\theta_{2}^{(i)}(t),\theta_{3}^{(i)}(t)\right) \\
        \mbf{0}_{3\times3} & \mbf{A}_\text{dss}^{(i)}(t) \ebm, 
\end{equation}
 and $\mbf{B}_{u}^{(i)}(t)$ is time-varying across the prediction horizon given the new expected magnetic field direction based on the orbital position and attitude at time $t$ in the $i^{th}$ iteration of the MPC problem. These continuous-time dynamics are converted to discrete-time in the same way as described in Section~\ref{sec:dyn_stateSpace} for use in the MPC policy.

The method for solving the LTV MPC policy using this linear approximation follows a simple iterative approach. Specifically, the following procedure is followed at each timestep $t$:
\begin{enumerate}
    \item An initial guess is found via the method of orbital propagation as in Section~\ref{sec:MPC_orb}.
    \begin{itemize}
        \item This solution defines the first iteration over the prediction horizon, denoted as inputs $\mathcal{U}_{t}^{(1)}$ and states $\mathcal{X}_{t}^{(1)}$.
    \end{itemize}
    \item The states from the previous iteration, $\mathcal{X}_{t}^{(i-1)}$, are used to solve for the time-varying dynamics matrix $\mbf{A}^{(i)}(t)$ as defined in Eq.~\eqref{eq:A_LTV} and the time-varying control input matrix $\mbf{B}_{u}^{(i)}(t)$ based on the expected magnetic field vector.
    \item The prediction model from Step 2 is used to solve the MPC problem in Eq.~\eqref{eq:MPC_prediction}.
    \item Steps 2 and 3 are repeated until both the magnetic field vector and satellite spin rates converge.
    \begin{itemize}
        \item Magnetic field convergence criteria:
        \begin{itemize}
            \item $\norm{\mbs{\Xi}^{(i-1)} - \mbs{\Xi}^{(i)}}_\infty < \xi_\text{diff,tol}$, where $\mbs{\Xi}^{(i)} \in \mathbb{R}^{N}$ is a matrix containing all magnetic field difference angles $\xi_{k}^{(i)}$ in the prediction horizon for the $i^\text{th}$ iteration and $\xi_\text{diff,tol}$ is the convergence tolerance.
            \item The difference angle is defined through the law of cosines as $\xi_{k}^{(i)} = \cos^{-1}\left(\frac{\mbf{b}_k^{(i-1)}\cdot\mbf{b}_k^{(i)}}{\norm{\mbf{b}_k^{(i-1)}}_2\norm{\mbf{b}_k^{(i)}}_2}\right)$ $\mbf{b}_k^{(i)}$ is the magnetic field vector at the $k^\text{th}$ timestep in the $i^\text{th}$ iteration.
        \end{itemize}
        \item Roll rate convergence criteria:
        \begin{itemize}
            \item $\norm{~|\mbs{\Gamma}^{(i-1)} - \mbs{\Gamma}^{(i)}|~}_\infty < \gamma_\text{diff,tol}$, where $\mbs{\Gamma}^{(i)} \in \mathbb{R}^{N+1}$ is a matrix containing all $\gamma_k^{(i)}$ in the prediction horizon for the $i^\text{th}$ iteration and $\gamma_\text{diff,tol}$ is the roll rate convergence tolerance.
        \end{itemize}
    \end{itemize}
    \item The optimal control input for the first timestep in the final iteration following successful convergence, $\mbf{u}_{0|t}^*$, is applied to the system, and the algorithm is repeated at Step 1 for the following timestep. 
\end{enumerate}

Figure~\ref{fig:orbLin_pred} shows the magnetic field and attitude errors resulting from this linear propagation method over a long prediction horizon. The intermediate solutions corresponding to each iteration for the linear propagation (`linprop') method are in gray, where each becomes darker than the last, and the orbital propagation (`orbProp') method is the first iteration solution for linProp. This result clearly illustrates a decrease in the magnetic field prediction error compared to the orbital propagation method due to the consideration of rotational motion. While the linear propagation method has a substantially lower magnetic field prediction error, the error in the attitude prediction sees no improvement. It is hypothesized that this is due to limitations in the linearization. 

\subsection{Nonlinear Propagation Prediction}\label{sec:MPC_nlin}

The final MPC policy makes use of the nonlinear attitude equations of motion to improve the accuracy of the prediction model. A true nonlinear MPC solution where a direct optimization of the nonlinear equations would far exceed computational capabilities of most spaceflight missions, especially for that of a CubeSat. Thus, this method instead linearizes about the state solution from the previous iteration via a single propagation of the nonlinear dynamics. This methodology is common in literature where sequential convex programming (SCP) methods are used~\cite{morgan2014}. The nonlinear propagation prediction method introduced in this section employs an idea of SCP where successive linearization enables model predictive accuracy through convergence of time-varying parameters. 

\begin{figure}[t!]
	\includegraphics[width=0.5\linewidth]{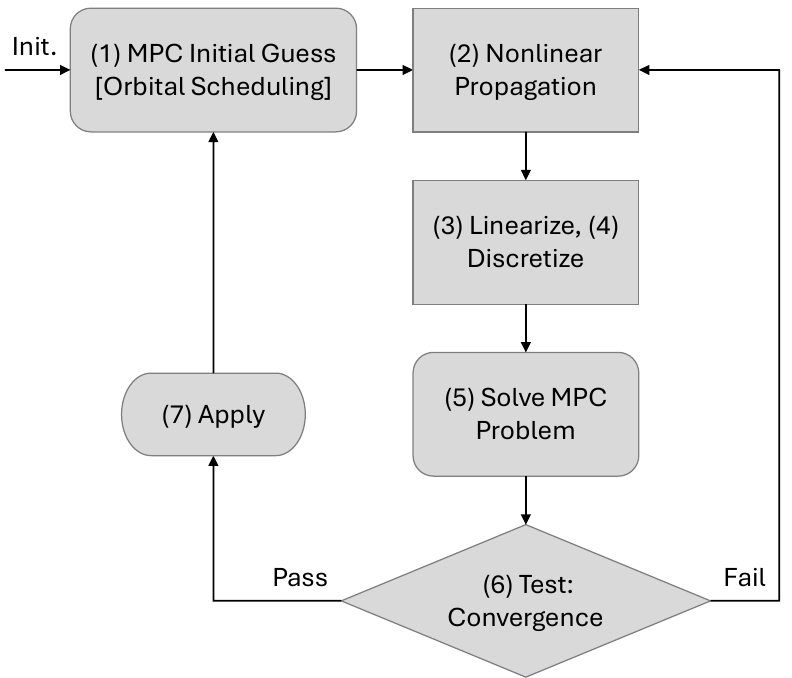} 
	\centering
	\caption{Flowchart for the nonlinear propagation MPC prediction method.}\label{fig:nLTV_flow}
\end{figure}

The method for solving the LTV MPC policy using a linearization about a nonlinear propagation follows an iterative approach similar to the method outlined in Section~\ref{sec:MPC_lin}. A flow diagram corresponding to this method is illustrated in Fig.~\ref{fig:nLTV_flow}, where the step numbering aligns with the algorithm, and the initialization (`init.') denotes the states at time $t$. Specifically, the following procedure is followed at each timestep $t$:
\begin{enumerate}
    \item An initial guess (inputs $\mathcal{U}_t^{(1)}$ and states $\mathcal{X}_t^{(1)}$) is found via the method of orbital propagation as in Section~\ref{sec:MPC_orb}.
    \item The control input from the previous iteration, $\mathcal{U}_t^{(i-1)}$, is used to propagate the nonlinear attitude dynamics over the prediction horizon in Eq.~\eqref{eq:DSS_full} assuming no external disturbances (i.e., $\mbs{\tau}_b^\text{ext} = \mbf{0}$).
    \begin{itemize}
        \item The nonlinear dynamics in Eq.~\eqref{eq:DSS_full} for this iteration have the form $\frac{\mathrm{d}}{\mathrm{d}t}\mbf{x}^{(i)}(t) = f\left(\mbf{x}^{(i)}(t),\mbf{u}^{(i)}(t)\right)$.
        \item This step provides the reference trajectory for the current iteration, $\bar{\mathcal{X}}_t^{(i)}, \bar{\mathcal{U}}_t^{(i)}$.
    \end{itemize}
    \item The reference trajectory from Step 2 is approximated with a first-order Taylor series linearization. Specifically,
    \begin{equation*}
        \frac{\mathrm{d}}{\mathrm{d}t}\mbf{x}^{(i)}(t) \approx \mbf{A}^{(i)}(t)\mbf{x}^{(i)}(t) + \mbf{B}_u^{(i)}(t)\mbf{u}^{(i)}(t) + \mbf{B}_w\hat{\mbf{w}}(t) + \mbf{z}^{(i)}(t),
    \end{equation*}
    where $\mbf{A}^{(i)}(t)$ and $\mbf{B}_u^{(i)}(t)$ are evaluated using the reference trajectory states $\bar{\mathcal{X}}_t^{(i)}$ with the same method as in Section~\ref{sec:MPC_lin} and $\mbf{z}^{(i)}(t) = f\left(\bar{\mbf{x}}^{(i)}(t),\bar{\mbf{u}}^{(i)}(t)\right) - \mbf{A}^{(i)}(t)\bar{\mbf{x}}^{(i)}(t) - \mbf{B}_u^{(i)}(t)\bar{\mbf{u}}^{(i)}(t)$. 
    \item The linear dynamics from Step 3 are discretized using a zeroth-order hold approximation.
    \item The prediction model from Step 4 is used to solve the MPC problem in Eq.~\eqref{eq:MPC_prediction}.
    \begin{itemize}
        \item It is crucial to note that this linearization recovers the true attitude states, $\mbs{\Theta}$ and $\mbs{\omega}_b^{ba}$, and that the constraints in the MPC problem in Eq.~\eqref{eq:Q2-MPC} should be modified accordingly. 
    \end{itemize}
    \item Steps 2-5 are repeated until both the magnetic field vector and satellite roll rate converge using the same criteria as in Section~\ref{sec:MPC_lin}.
    \item The optimal control input for the first timestep in the final iteration, $\mbf{u}_{0|t}^*$, is applied to the system, and the algorithm is repeated at Step 1 for the following timestep. 
\end{enumerate}

\begin{figure}[t!]
	\centering
	\begin{subfigure}[]{0.4\textwidth}
		\includegraphics[scale=0.6]{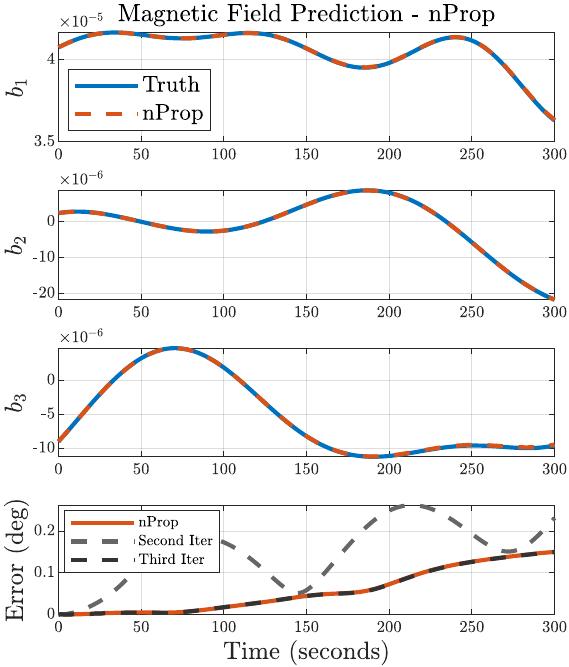}
        \caption{Magnetic field vector components}
        \centering
	\end{subfigure}%
	\begin{subfigure}[]{0.6\textwidth}
		\includegraphics[scale=0.6]{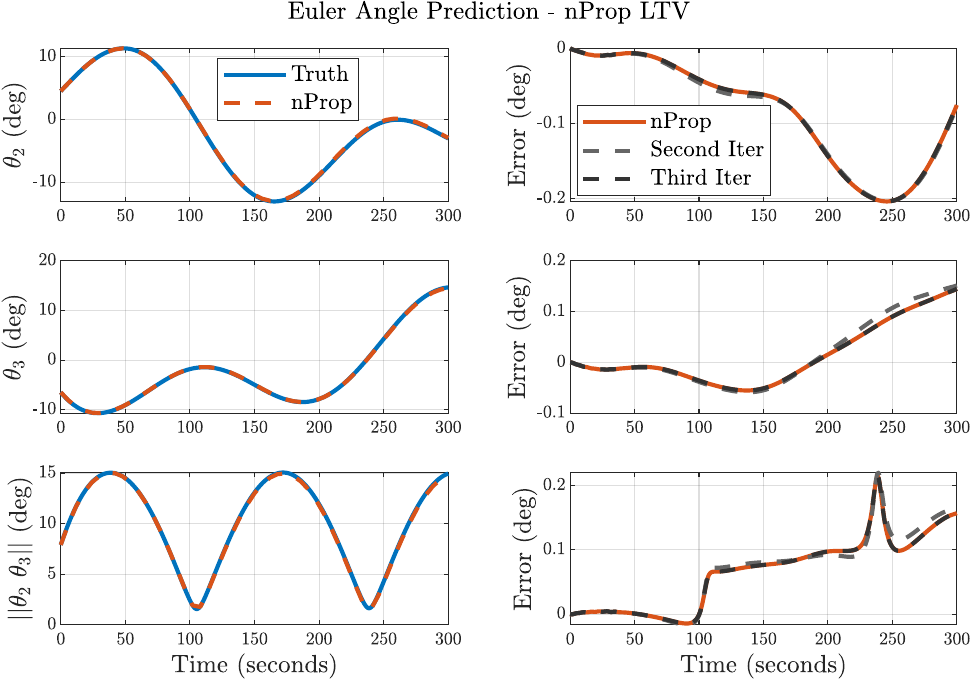}
        \caption{Pitch-yaw, and norm values and errors}
        \centering
	\end{subfigure}%
	\centering
	\vspace{-8pt}
	\caption{Nonlinear propagation prediction method expected values.}\label{fig:nLin_pred}
\end{figure}

Figure~\ref{fig:nLin_pred} shows the improvement made in the attitude and magnetic field prediction error when using the successive linearization method via nonlinear propagation (`nProp').  Intermediate solutions for each iteration are included in this figure, but the first iteration is omitted for clarity, as this is the solution to the propagation method described in Section~\ref{sec:MPC_orb}. The pointing norm error and maximum magnetic field prediction errors both are maintained below $0.2^\circ$ for the overwhelming majority of the long prediction horizon. This improvement in prediction quality is expected to correspond to improvements with regards to satellite performance and robustness to model uncertainty. 

\section{Simulation Results}\label{sec:results}

Many CubeSats are deployed from the International Space Station (ISS), thus, numerical simulations are performed with the satellite placed in a circular, near-ISS orbit at an altitude of $420$~km with an inclination of $50^\circ$. The orbit is propagated through Keplerian equations of motion including the J2 perturbation. See Ref.~\cite{DeRuiter2013} for more details on the orbital equations of motion and the disturbances that can affect the evolution of the satellite's orbit. For all simulations, the 2020 World Magnetic Model is used to simulate the Earth's magnetic field~\cite{chulliat2020us}.

The CubeSat takes the form of a 3U (i.e., dimensions of $30\times10\times10$~cm) in a solar-panel-deployed configuration with a moment of inertia matrix $\mbf{I}_b = \text{diag}\{1,2,2\}\cdot 10^{-2}$~kg/m$^2$. It is assumed to have drag coefficient $C_d = 2.5$, and the simulation is performed under high solar activity where the air density is $\rho = 4.02\times 10^{-11}$~kg/m$^3$. The nominal roll rate is $\omega_{b1}^{ba} = \gamma_\text{nom} = 0.75^\circ$/s, which is the value used in the linearized dynamics in Eq.~\eqref{eq:CT_LTV}. The satellite is equipped with a momentum wheel that has an inertia of $2$~kg$\cdot$mm$^2$ that is pointed in the direction of the boresight axis and is nominally spinning at a rate of $400$~rad/s. The residual magnetic dipole for the disturbance torque $\mbs{\tau}_b^\text{mag,d}$ is chosen as $\mbf{m}_d^\trans = \left[ 0.1 \,\,\, -0.1 \,\,\, 0.15 \right]\times 10^{-4}$~A$\cdot$m$^2$. 

All simulations presented in this paper utilize the same constraints on control input saturation and system states. Each magnetic torque rod has a saturation limit of $u_{m,\text{max}} = 0.48$~A$\cdot$m$^2$, whereas the maximum angular acceleration of the reaction wheel is set to $u_{w,\text{max}} = 10~\text{rad/sec$^2$}$. The hard constraint on the minimum allowable roll rate is chosen to be $\gamma_\text{min} = 0.05^\circ/\text{sec}$, while the soft constraint on the minimum roll rate is $\gamma_\text{min,s} = 0.25^\circ/\text{sec}$. The maximum roll rate is $\gamma_\text{max,s} = 1.5^\circ/\text{sec}$ to ensure the satellite does not begin to spin too fast for communication or other ADCS systems to be able to perform as desired. The allowable pointing half-cone angle soft constraint as defined in Eq.~\eqref{eq:cone_constraint} is $\alpha_\text{max,s} = 15^\circ$. 

Although it is essential that a given control policy ensures that the spacecraft attitude does not drift beyond the prescribed constraints, it is also desirable for the control policy to minimize the control input. This paper quantifies magnetic torque rod use in two ways: one as a time history of the control torque applied to the satellite from the magnetic actuation, and the other as the total torque rod input applied over a simulation as a means of comparing the control effort required across initial conditions and MPC policy types. The control input is applied with a zero-order hold, therefore the total torque rod input is calculated at each time step and summed over the entire simulation, such that
\begin{equation}\label{eq:tr_tot}
    u_{m,\text{tot}} = \sum_{j=1}^n \left[\left(|m_{1,j}| + |m_{2,j}| + |m_{3,j}|\right)\Delta t\right],
\end{equation}
where $n$ is the total number of time steps in the simulation. When comparing results across simulations with different lengths in time, a normalization with respect to time is required such that $u_{m,\text{int}} = \frac{1}{n\Delta t}u_{m,\text{tot}}$.

The MPC policies are applied in feedback to a simulation of the spacecraft and the LEO environment. This simulation follows the nonlinear attitude and translational equations of motion complete with disturbances as described in Section~\ref{sec:dynamics}. In this way, the MPC prediction approaches are tested in a reasonably high-fidelity simulation that emulates the differences to be expected between the MPC prediction model and the actual orbital environment in practice. 

\subsection{Warm-Starting the Quadratic Program}
Each simulation herein employs a simple form of `warm-starting' the MPC problem after the initial timestep. In a general sense, these methods ease the computational burden on the receding-horizon MPC policy by using the sequence of future control inputs solved at the previous timestep $(t-\Delta t)$ to inform a rough estimate of the optimal solution to the next time the MPC optimization problem is solved~\cite{otta2015}. In this paper, the method of warm-starting aims to assist with convergence of the LTV properties when using the algorithms in Sections~\ref{sec:MPC_lin} and~\ref{sec:MPC_nlin}. The states from the solution at the previous timestep are used to inform the time-varying parameters in the state dynamics and control matrices over the entire prediction horizon, and the last timestep assumes the same states as the previous (i.e., $\mbf{x}_{N-1|t} = \mbf{x}_{N|t}$).  

The QCQP MPC problem in Eq.~\eqref{eq:Q2-MPC} is solved in MATLAB using the linear matrix-inequality parser YALMIP~\cite{yalmip} with mosek~\cite{mosek} as the solver. Whenever nonlinear propagation is needed, such as for the algorithm in Section~\ref{sec:MPC_nlin}, a variable-timestep ordinary differential equation integrator, ODE45, is used to propagate the expected nonlinear dynamics over the prediction horizon.

\subsection{MPC Policy Tuning}

An important component of the quadratic cost function in Eq.~\eqref{eq:Q2-MPC} are the weights that regulate the states and control inputs over the prediction horizon. The objective of this cost function is to encourage state drift within the constraint set while minimizing control effort. To achieve this, the state weighting matrix is chosen to place the greatest relative importance on tracking the roll rate of the CubeSat and on regulating the pitch and yaw angles, such that $\mbf{Q} = \frac{\Delta t\times 10^{-3}}{7.5}\cdot\text{diag}\{10^{-12},1,1,10,10^{-2},10^{-2}\}$. The control weighting matrix is $\mbf{R} = \frac{7.5\times10^5}{\Delta t}\cdot\text{diag}\{10,1,1,1\}$ to penalize reaction wheel input such that the MPC policy largely relies on the magnetic actuation. The exact procedure to find these weights is omitted for brevity, but largely follows intuition with some iteration through simulation testing.

Soft constraint cost weights are selected to allow for some state deviation from the constraint bound, while penalizing any significant deviation. The weights for the roll rate soft constraints are chosen as $\psi_1 = \psi_2 = 10^4$, and the weight for the pointing cone constraint is $\psi_3 = 10^5$ to have a greater comparative importance on minimizing drift from the allowable pointing cone. 

\begin{figure}[t!]
	\centering
	\begin{subfigure}[]{0.5\textwidth}
		\includegraphics[scale=0.585]{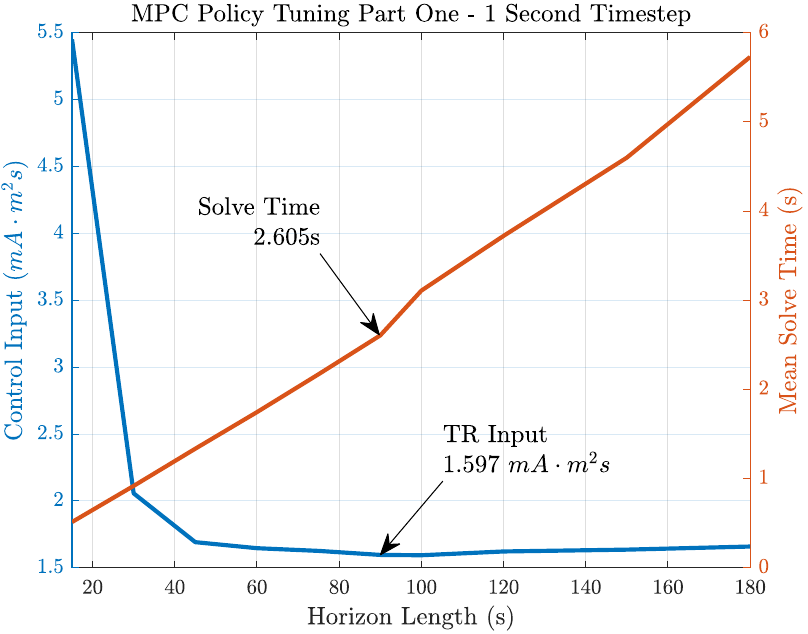}
        \caption{Fixed timestep}\label{fig:tuning_p1}
        \centering
	\end{subfigure}%
	\begin{subfigure}[]{0.5\textwidth}
		\includegraphics[scale=0.585]{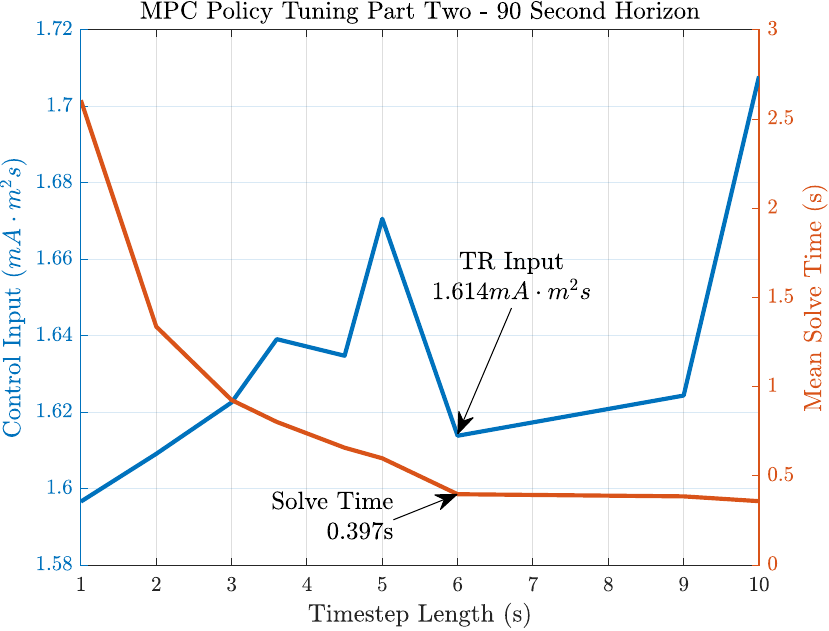} 
        \caption{Variable timestep}\label{fig:tuning_p2}
        \centering
	\end{subfigure}%
	\centering
	\vspace{-8pt}
	\caption{Prediction horizon tuning procedure steps 1 and 2. Note that y-axis varies between the sub-figures.}\label{fig:MPC_tuning}
\end{figure}

The central component of any MPC policy is the prediction horizon length, which is defined via the length of the discretization timestep, $\Delta t$, and the number of timesteps in the prediction, $N$. In a general sense, a longer total prediction horizon, $N\Delta t$, is expected to result in improved performance with diminishing returns with regards to computation time. Meanwhile, the accuracy of the discretization used in the prediction horizon is expected to be greater with a small $\Delta t$, but this would require more timesteps to cover the desired overall prediction horizon length. This tradeoff is illustrated in two parts of tuning the MPC policy in Fig.~\ref{fig:MPC_tuning}. The simulations for this procedure use the nonlinear propagation MPC policy described in Section~\ref{sec:MPC_nlin} with the same initial conditions as in the open-loop simulations in Section~\ref{sec:results_OL} over 1/2 of an orbit. Part one of the tuning process uses a fixed timestep $\Delta t = 1$~second and varies the horizon length via increasing $N$. The direct tradeoff between performance and computation time is evident with the consistent increase in the average time-to-solve in Fig.~\ref{fig:tuning_p1}. The control input decreases drastically as the prediction horizon increases, thus a prediction horizon length is chosen to be 90 seconds.

In part two, this horizon length is held and the timestep length is increased while modifying $N$ accordingly. It is generally expected that as the timestep length increases, computation time will greatly improve while performance will slightly degrade as the discretization is less accurate. This expectation is confirmed in Fig.~\ref{fig:tuning_p2} as the computation time decreases significantly as the timestep length increases. The selected timestep length from this tuning procedure is $\Delta t = 6$~seconds with a horizon length of $N=15$~timesteps to provide a meaningfully low solve time without much loss in control performance. The results from this prediction horizon tuning procedure are used in definitions of $\Delta t$ and $N$ for each simulation herein.

\subsection{Controlled Two-Orbit Simulations with Initial Drift}

\begin{figure}[t!]
	\centering
	\begin{subfigure}[]{0.5\textwidth}
		\includegraphics[scale=0.63]{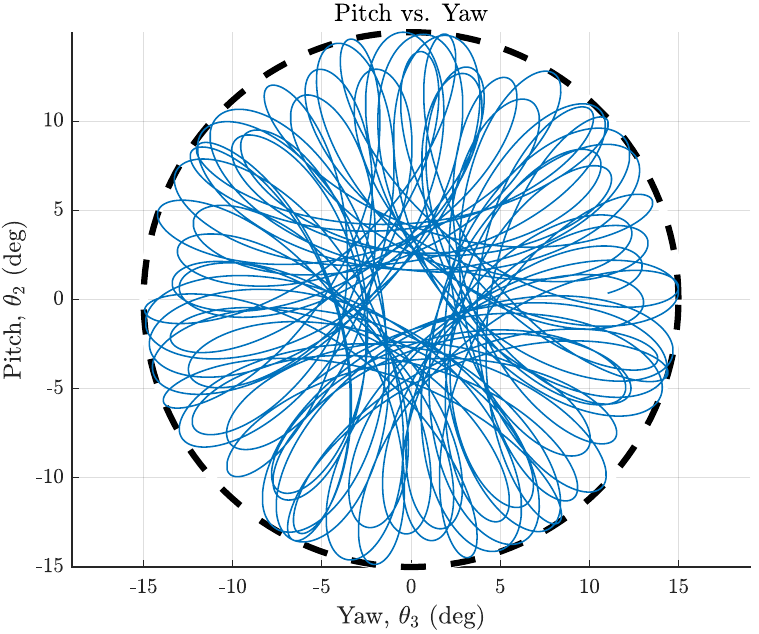}
        \caption{Pitch-yaw trajectory}\label{fig:2orb_nPropA}
        \centering
	\end{subfigure}%
	\begin{subfigure}[]{0.5\textwidth}
		\includegraphics[scale=0.63]{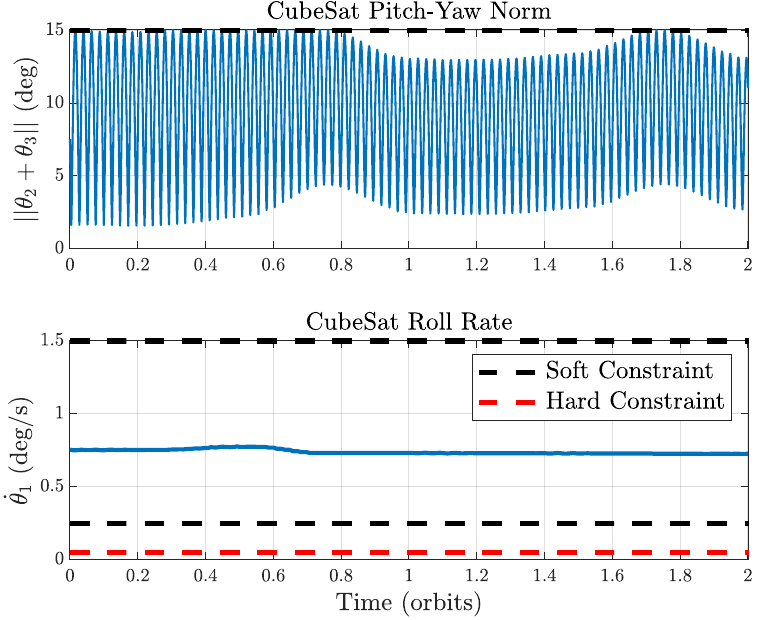} 
        \caption{Pitch-yaw norm and roll rate}
        \centering
	\end{subfigure} \\
	\begin{subfigure}[]{0.5\textwidth}
		\includegraphics[scale=0.63]{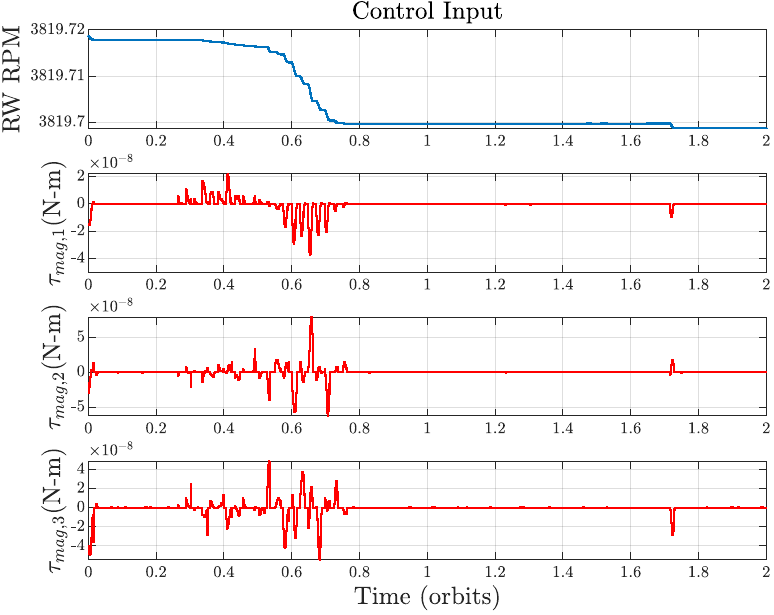} 
        \caption{Control input}
        \centering
	\end{subfigure}%
	\centering
	\vspace{-8pt}
	\caption{2-orbit simulation results using the nonlinear propagation LTV MPC policy.}\label{fig:2orb_nProp}
\end{figure}

A series of two-orbit simulations are completed across varying initial conditions to compare the three LTV MPC policies described in Section~\ref{sec:MPC}. Twenty different initial states are tested, each ensuring the system reaches the half-cone pointing angle constraint at least once during the course of the simulation. An example case is shown in Fig.~\ref{fig:2orb_nProp} using the nonlinear propagation MPC policy, where the initial attitude is $\mbs{\Theta}(0) = \left[ 0 \,\, -4.858 \,\, -5.757 \right]^\trans$~degrees with angular velocity $\mbs{\omega}_b^{ba}(0) = \left[ (\gamma+4.584\times10^{-3}) \,\,\, 0.272 \,\,\, 0.169 \right]^\trans$~degrees/sec. Figure~\ref{fig:2orb_nPropA} shows the pitch vs. yaw state trajectory over time, with the pointing cone soft constraint depicted as a black dashed line. These results demonstrate the capability for the MPC problem to maintain the constrained states within their allowable regions via the small torques from the magnetic actuation and a negligible reaction wheel RPM variance. This is a compelling example that demonstrates the policy's capability to react to poor initial conditions immediately, even in the face of external disturbances and limited actuation magnitude availability.

\begin{figure}[t!]
	\centering
	\begin{subfigure}[]{0.5\textwidth}
		\includegraphics[scale=0.63]{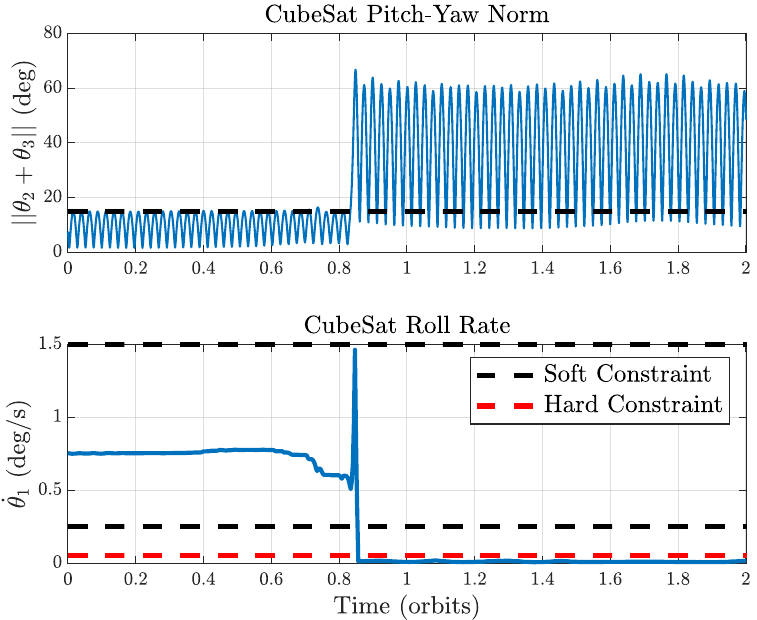}
        \caption{Pitch-yaw norm and roll rate}
        \centering
	\end{subfigure}%
	\begin{subfigure}[]{0.5\textwidth}
		\includegraphics[scale=0.63]{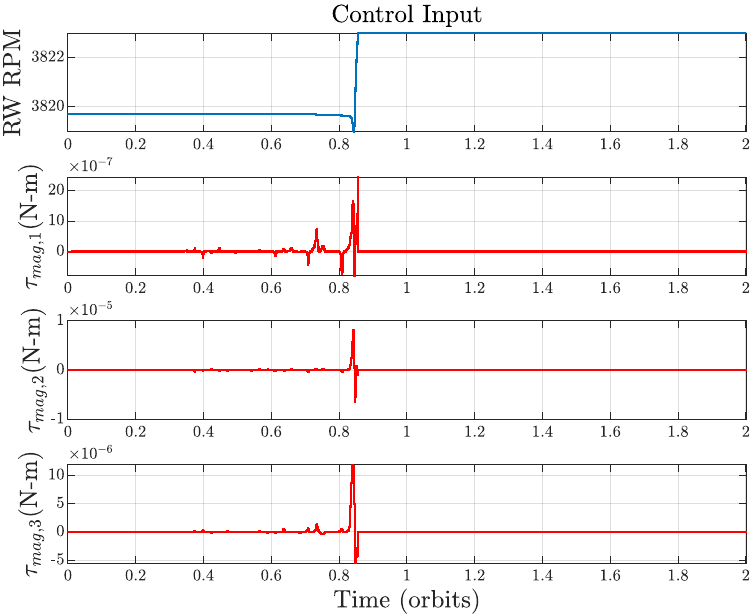} 
        \caption{Control input}
        \centering
	\end{subfigure}%
	\centering
	\vspace{-8pt}
	\caption{2-orbit simulation results using the linear propagation LTV MPC policy with a failure past 0.8 orbits.}\label{fig:2orb_linProp}
\end{figure}

Figure~\ref{fig:2orb_linProp} illustrates a case where an imperfect model is used in the prediction---specifically the linear propagation MPC policy. It is clear that the constraint set and limited actuation availability coupled with the model inaccuracy in the prediction caused the system to reach a region of infeasibility past the minimum allowable roll rate and never recover. This result is a testament to the capabilities of the nonlinear propagation policy, as it is able to accurately predict the future response and act accordingly in the present, whereas the linear propagation policy is forced to react to unexpected soft constraint violation and does not have the control authority required to be able to respond quickly. The response using the orbital propagation policy is similar to the nonlinear propagation policy for these initial conditions, and results are omitted for brevity. 

\subsubsection{Performance Comparison}

\begin{figure}[t!]
	\centering
	\begin{subfigure}[]{0.5\textwidth}
		\includegraphics[scale=0.63]{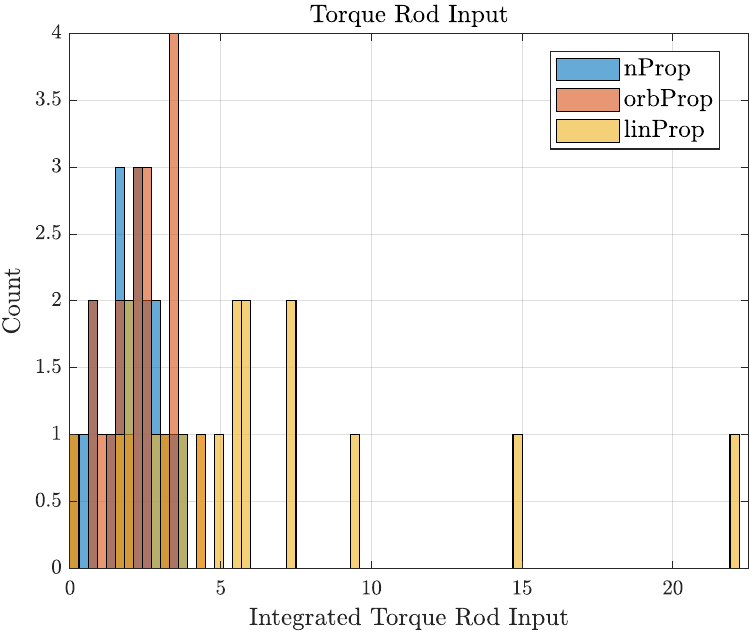}
        \caption{All LTV MPC policies}\label{fig:perfComparisonA}
        \centering
	\end{subfigure}%
	\begin{subfigure}[]{0.5\textwidth}
		\includegraphics[scale=0.63]{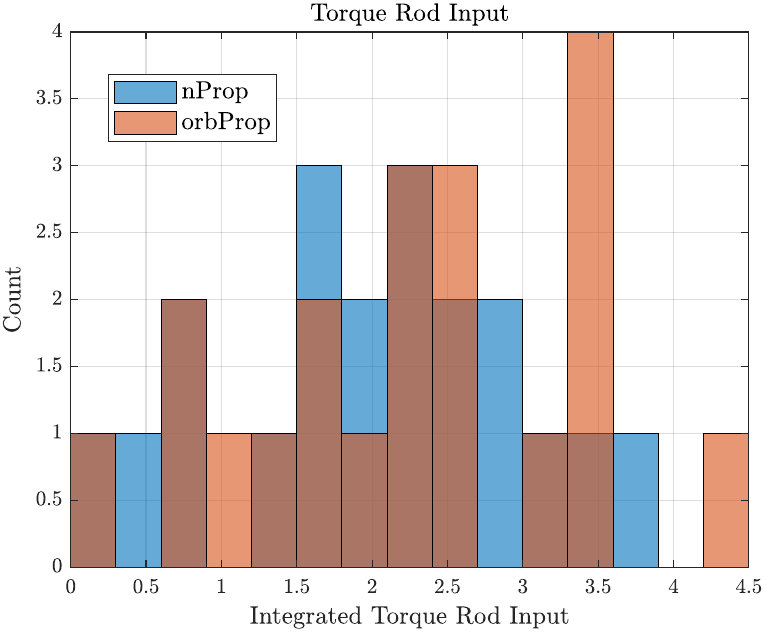} 
        \caption{Linear propagation excluded}\label{fig:perfComparisonB}
        \centering
	\end{subfigure}%
	\centering
	\vspace{-8pt}
	\caption{Torque rod input histograms from completed short simulations.}\label{fig:perfComparison}
\end{figure}

The performance with respect to torque rod input for each of the three MPC policies is compared in the histograms of Fig.~\ref{fig:perfComparison}. It is important to note that two simulations using the linear propagation policy failed due to reaching a region of infeasibility, and thus only data from the eighteen simulations for this policy that concluded the simulation are included. It is immediately clear that the nonlinear and orbital propagation policies far outperform the linear propagation, thus Fig.~\ref{fig:perfComparisonB} is included to more clearly compare the torque rod input required of these two policies. The nonlinear propagation policy has slightly better performance overall compared to the orbital propagation method, which demonstrates the appeal in using a more accurate prediction model beyond the robustness improvement. 

To compare these MPC policies between the specific initial conditions, Table~\ref{table:performance} shows the number of times the three prediction types had the lowest torque rod input used for each specific set of initial conditions. The nonlinear propagation policy massively outperforms the other policies via this metric. Further, for the six initial conditions where the nonlinear propagation did not have the best performance, the increase in torque rod input was only an average of 4.49\% from that of the orbital propagation. Regarding the fourteen simulations where the orbital propagation did not have the lowest torque rod input, the increase in the input relative to the nonlinear propagation was 17.05\%. This clearly demonstrates that while the nonlinear propagation policy may not always perform the best in these short simulations, the relative increase in torque rod input required is much smaller compared to the other policies. Importantly, none of the nonlinear or orbital propagation prediction policies failed before completing the two orbits. 

\begin{table}
    \centering
    \caption{MPC policy short simulation comparison}
    \vspace{-8pt}
    \begin{tabular}{c | c | c}
    \hline \hline
        Prediction Type & \# of `Best' Runs & Failed Runs \\
        \hline\hline
        Orbital Propagation & 6 & 0 \\
        \hline
        Linear Propagation & 0 & 2 \\
        \hline
        Nonlinear Propagation & 14 & 0 \\
        \hline
    \end{tabular}
    \label{table:performance}
\end{table}

\begin{figure}[t!]
	\centering
	\begin{subfigure}[]{0.5\textwidth}
		\includegraphics[scale=0.63]{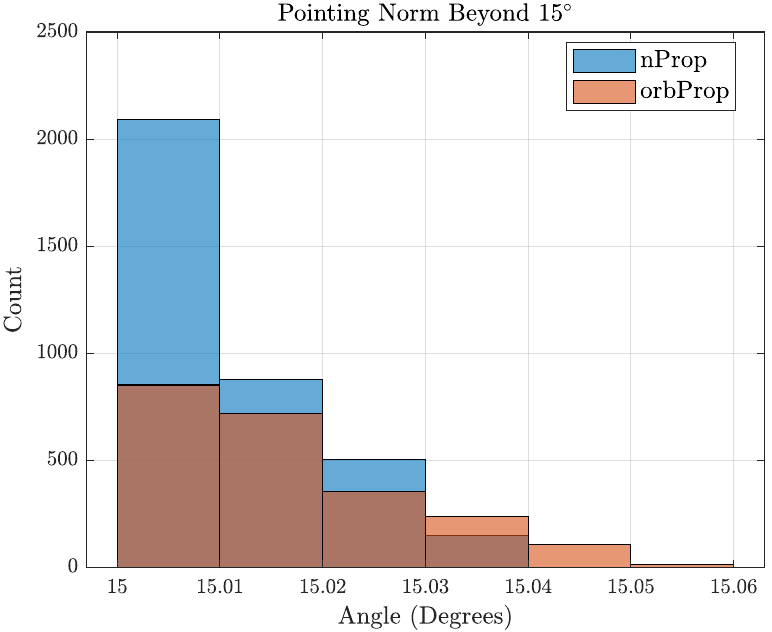}
        \caption{Nonlinear and orbital propagation policies}\label{fig:errorComparisonA}
        \centering
	\end{subfigure}%
	\begin{subfigure}[]{0.5\textwidth}
		\includegraphics[scale=0.63]{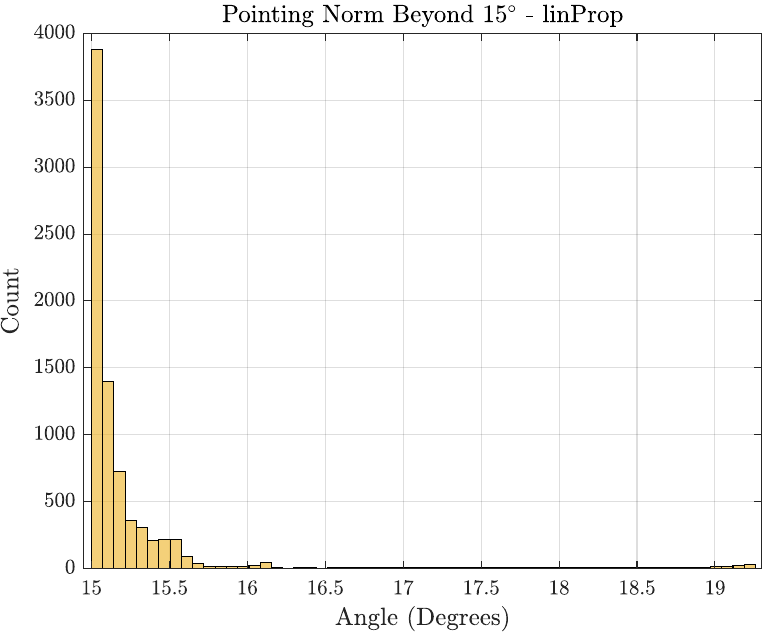} 
        \caption{Linear propagation policy}\label{fig:errorComparisonB}
        \centering
	\end{subfigure}%
	\centering
	\vspace{-8pt}
	\caption{Pointing cone soft constraint violation histograms.}\label{fig:errorComparison}
\end{figure}

Another important metric of performance is the amount of time each policy violates the soft constraints; especially the pointing cone constraint. Figure~\ref{fig:errorComparison} includes histograms showing the number of times each policy violated this constraint. Note that the count does not denote the discrete timesteps used in the MPC calculation, but instead are the number of instances where the constraint is violated in the nonlinear simulation that is used to emulate the vehicle and environmental dynamics, saved at a rate of 5Hz.  
The linear propagation error in Fig.~\ref{fig:errorComparisonB} is clearly the worst performer with regards to pointing norm constraint violation. Meanwhile, the nonlinear propagation sees several small ($<0.01^\circ$) deviations, far exceeding that of the orbital propagation. This is assumed to be due to the more accurate model moving closer to the constraint boundary to maximize drift, where small deviations in the prediction model vs. the truth cause the system to deviate slightly from the soft constraint. The nonlinear propagation sees no constraint violation greater than $0.04^\circ$, while the orbital propagation experiences several. This again demonstrates the importance of using an accurate model, where the allowable constraint set will be used as much as possible by the predictive control policy to minimize the input cost, while not deviating too far from the soft constraint boundaries. 

\subsubsection{Computational Burden}\label{sec:results_comp}

\begin{figure}[t!]
	\includegraphics[width=0.55\linewidth]{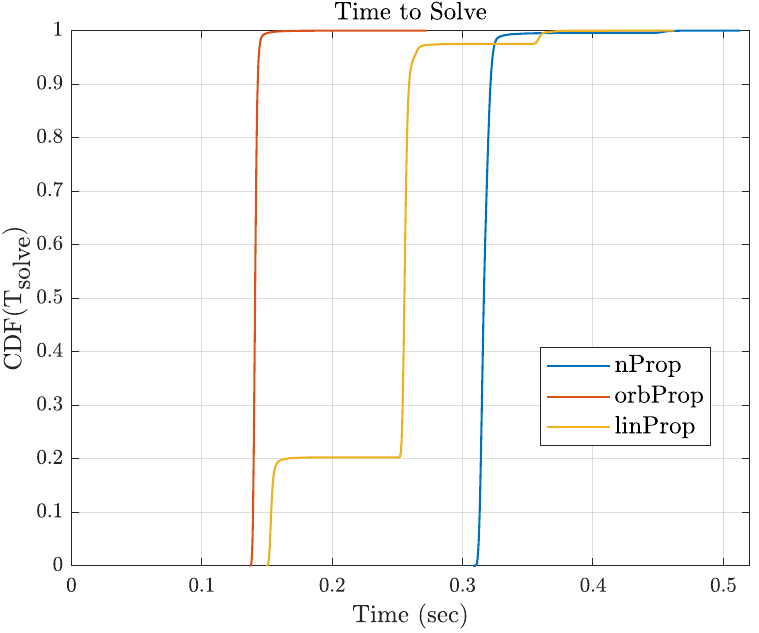} 
	\centering
	\caption{MPC policy solve time cumulative distribution for each prediction method.}\label{fig:cdfComp}
\end{figure}

Each simulation is performed on a desktop computer equipped with an AMD Ryzen 5 5600x processor. The solve times provided in this section aim to provide a rough measure of comparison and feasibility, rather than a rigorous assessment of the true computation time that would be reasonable on a CubeSat computing system. Figure~\ref{fig:cdfComp} includes data on the solve time for each MPC policy across the twenty simulations in the form of an empirical cumulative distribution function plot. This includes the entire time used to solve the MPC policy at each timestep according to the respective LTV prediction method. It is expected that in practice the MPC policy would be solved directly as a SOCP, thus the time to parse the problem is removed from the data for each iteration at each timestep. Key takeaways from this data are listed in Table~\ref{table:computation} to show important values of computation time, where the values given for each method is associated with a percentile (e.g., 99\% of the solve times for the nonlinear propagation method take less than 0.33 seconds). 

The linear and nonlinear propagation LTV policies have greater computational burden than that of the orbital propagation policy due to the iterative nature to solve the LTV problems. The effect of a higher iteration count is shown in the solve time for the linear propagation policy, where a distinctive bi-modal distribution is evident. Meanwhile, the nonlinear propagation method demonstrates a very consistent solve time of just over 0.3 seconds due to one iteration being required for each timestep. 

\begin{table}[t!]
    \centering
    \caption{Solve times associated with each percentile for the LTV MPC policies}
    \vspace{-8pt}
    \begin{tabular}{c | c | c | c}
        \hline\hline
     & \multicolumn{3}{c}{Computation Time (sec)} \\
        \hline
        Percentile & nProp & orbProp & linProp \\
        \hline\hline
        100\% & 0.513 & 0.272 & 0.462 \\
        \hline
        99.73\% & 0.453 & 0.154 & 0.365 \\
        \hline
        99\% & 0.330 & 0.146 & 0.359 \\
        \hline
        95.4\% & 0.323 & 0.144 & 0.264\\
        \hline
    \end{tabular}
    \label{table:computation}
\end{table}

There are many potential ways in which the computational burden of each MPC policy outlined in this paper could be reduced to address the limited computational capabilities on board space vehicle processing units. It is expected that the computational efficiency of the SOCPs solved in this paper could be greatly improved when written in languages commonly used in space vehicle operations (such as C) rather than the use of MATLAB and 3rd-party toolboxes. Further, significant improvement in solve time for the nonlinear propagation policy could be achieved by using a Runge-Kutta integration method written specifically for this problem in the satellite's flight hardware rather than implementing an algorithm similar to MATLAB's ODE45 as was used in this work.

\subsection{Mission-Duration Simulations}

Long simulations are performed to demonstrate the capability of the nonlinear propagation and orbital scheduling MPC policies to handle cumulative drift due to disturbances rather than simply the effect of initial conditions. Thus, the initial states for the results in this section are $\mbs{\Theta}(0) = \left[ 0 \,\, 1 \,\, -0.5 \right]^\trans$~degrees and $\mbs{\omega}_b^{ba}(0) = \left[ \gamma \,\, 0.01 \,\, -0.075 \right]^\trans$~degrees/sec. Due to disturbances, the system reaches the half-cone angle constraint of $15^\circ$~degrees in 18 orbits with these initial conditions.  The linear propagation policy is not included in this section given the tendency to fail in the two-orbit simulations. 

\begin{figure}[t!]
	\centering
	\begin{subfigure}[]{0.5\textwidth}
		\includegraphics[scale=0.63]{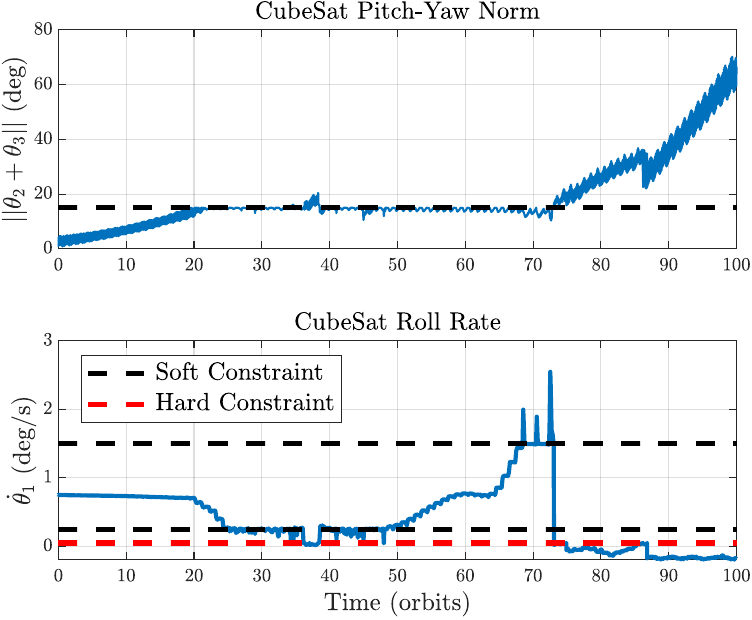}
        \caption{Pitch-yaw norm and roll rate}
        \centering
	\end{subfigure}%
	\begin{subfigure}[]{0.5\textwidth}
		\includegraphics[scale=0.32]{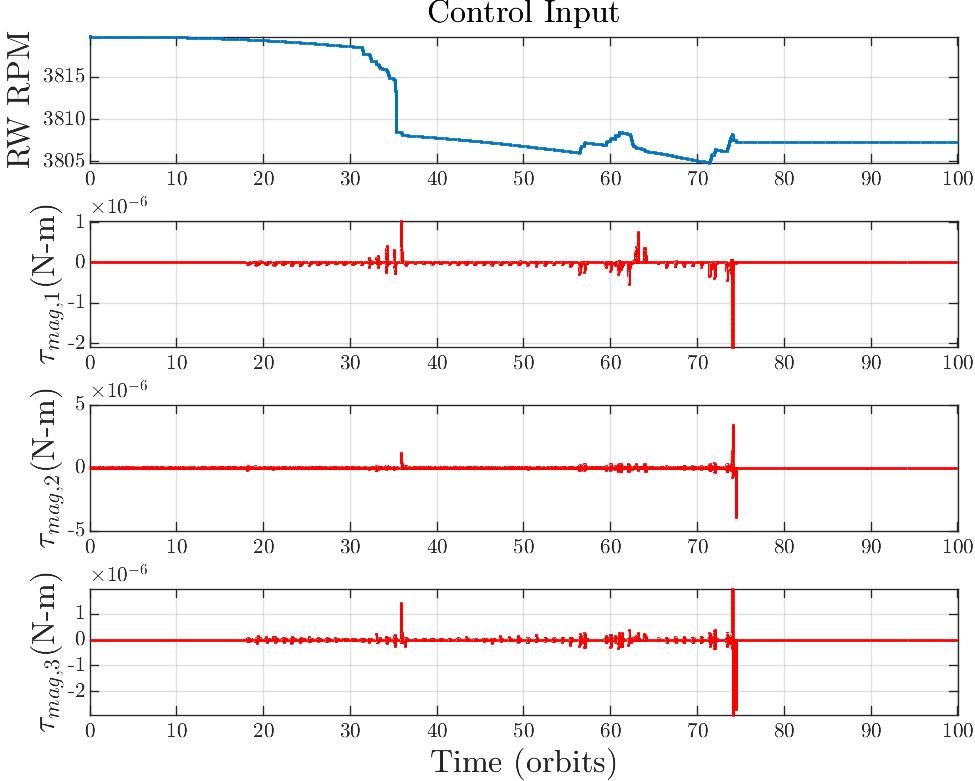} 
        \caption{Control input}
        \centering
	\end{subfigure}%
	\centering
	\vspace{-8pt}
	\caption{Failed 100-orbit simulation orbital propagation policy results.}\label{fig:orbLong}
\end{figure}

Figure~\ref{fig:orbLong} shows the response of the orbital scheduling MPC policy over a 100-orbit simulation. At approximately orbit 68, the roll rate maximum soft constraint is violated significantly, and the system never recovers until reaching a region of infeasibility in the optimization problem by violating the minimum roll rate constraint near orbit 73. Further, this simulation uses an optimization problem where the cost function is augmented with a terminal cost, $\mbf{x}_{k|t}^\trans\mbf{P}\mbf{x}_{k|t}$, where $\mbf{P}$ is the solution to the discrete algebraic Riccati equation (DARE) given the expected terminal state. This is added in an attempt to improve closed-loop stability, as simulations that performed using the nominal cost function in Eq.~\eqref{eq:Q2-MPC} failed in a similar way.

\begin{figure}[t!]
	\centering
	\begin{subfigure}[]{0.5\textwidth}
		\includegraphics[scale=0.32]{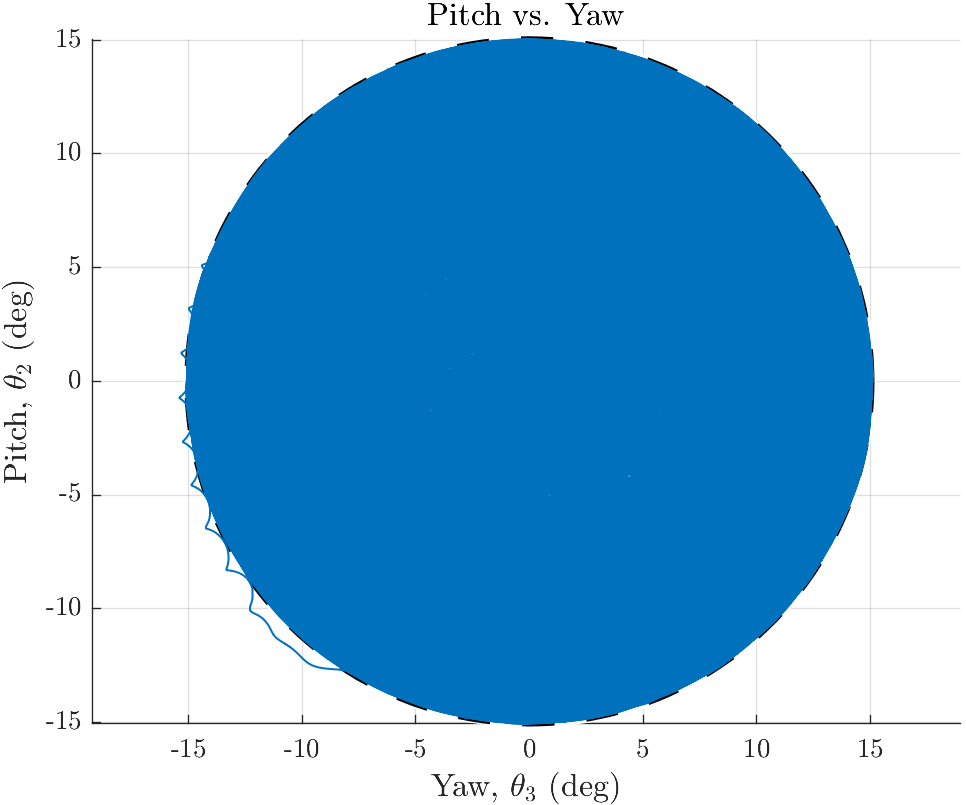}
        \caption{Pitch-yaw trajectory}
        \centering
	\end{subfigure}%
	\begin{subfigure}[]{0.5\textwidth}
		\includegraphics[scale=0.63]{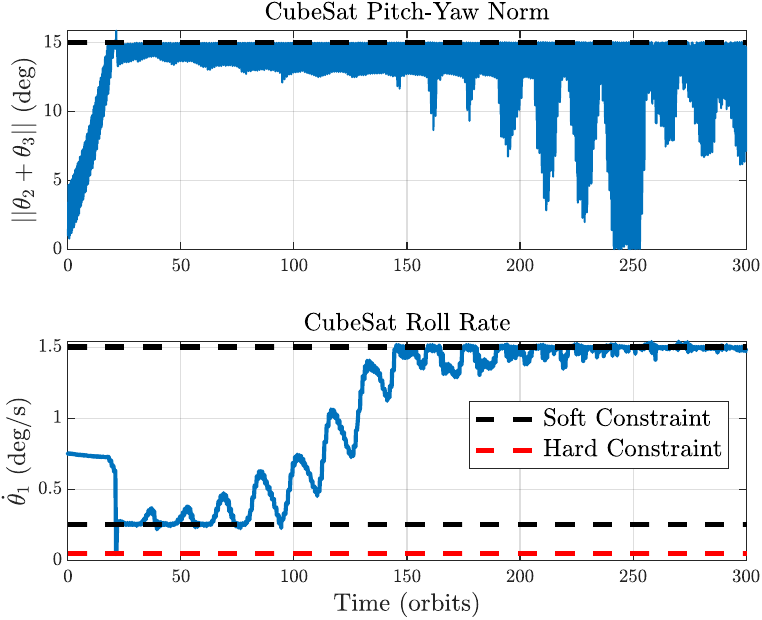} 
        \caption{Pitch-yaw norm and roll rate}
        \centering
	\end{subfigure} \\
	\begin{subfigure}[]{0.5\textwidth}
		\includegraphics[scale=0.32]{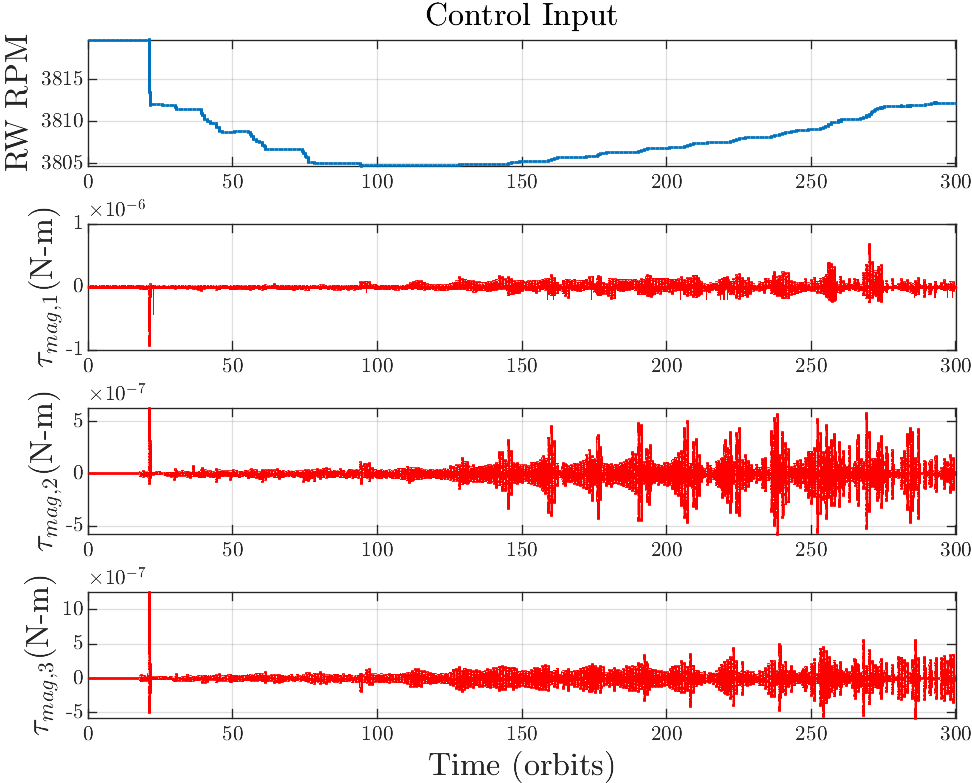} 
        \caption{Control input}
        \centering
	\end{subfigure}%
	\centering
	\vspace{-8pt}
	\caption{Successful 300-orbit simulation nonlinear propagation policy results.}\label{fig:nLTV_long}
\end{figure}

Conversely to the failure using the orbital scheduling policy, the nonlinear propagation policy is applied to the system using the same initial conditions over 300 orbits, with results shown in Fig.~\ref{fig:nLTV_long}. Given the accurate prediction model used in this MPC policy, the optimization problem is capable of responding to expected disturbances before any hard constraints are violated. The 300-orbit simulation equates to 19.3 days in the specific orbit, which is a substantial portion of science operations for typical CubeSat missions. The system deviates briefly from the pointing cone and minimum roll rate soft constraints early on, but recovers well and continues to perform as expected for the complete duration of the simulation.

\section{Conclusions}

This paper demonstrated the efficacy in using a time-varying predictive control policy for a dual-spin stabilized CubeSat that is primarily actuated with magnetic torque rods. As an extension to Ref.~\cite{halverson2024}, three different time-varying prediction models were incorporated within an MPC framework, each showing improvement in the accuracy of the important predicted states and magnetic field vector relative to an LTI prediction model. Several simulations were performed to demonstrate the capability for the constrained LTV MPC policies to respond to challenging initial conditions and environmental disturbances. It was found that the nonlinear-propagated LTV MPC policy outperformed the orbital scheduling and linearly-propagated policies in most metrics of control input and state error, and importantly never failed due to reaching a region of infeasibility in the optimization problem---a testament to the accuracy of the prediction model. There is a natural tradeoff with regards to computational resources required, however the nonlinear-propagated MPC policy had reasonable computation time for a preliminary investigation in computation time, and potential methods for mitigating this were presented.

Future work will include implementing methods to decrease the computational burden as described in Section~\ref{sec:results_comp}. Further, more realistic effects will be included in the nonlinear simulation and the MPC optimization problem, such as minimum actuation limits on torque rods and other non-convex constraints such as Sun keep-out zones. The operational regimes of this problem will also be further explored, such as the control policy used to take the space vehicle from de-tumble to the dual-spin configuration, as well as scheduled slews for ground station communication. Ultimately, the methods presented in this paper are intended to be included in an actual satellite mission, and to achieve this, these effects will need to be taken into account. 

\section*{Acknowledgments}

R. D. Halverson acknowledges support by the Science, Mathematics, Mathematics, and Research for Transformation (SMART) Scholarship-for-Service Program within the Department of Defense, USA. The authors would also like to acknowledge Professor Demoz Gebre-Egziabher with the University of Minnesota - Twin Cities, along with the University of Minnesota Small Satellite Research Laboratory, for their assistance in the conceptual aspects of this work. 

\bibliography{bibliography}

\end{document}